\documentclass[%
reprint,
superscriptaddress,
showpacs,preprintnumbers,
verbatim,
amsmath,amssymb,
aps,
pre,
]{revtex4-1}

\usepackage{color}
\usepackage{graphicx}
\usepackage{dcolumn}
\usepackage{bm}

\begin {document}

\title{%
  Langevin equation with fluctuating diffusivity: a two-state model
}

\author{Tomoshige Miyaguchi}
\email{tmiyaguchi@naruto-u.ac.jp}
\affiliation{%
  Department of Mathematics Education, 
  Naruto University of Education, Tokushima 772-8502, Japan}

\author{Takuma Akimoto}
\affiliation{%
  Department of Mechanical Engineering, Keio University, Yokohama, 223-8522, Japan
}%

\author{Eiji Yamamoto}
\affiliation{%
  Department of Mechanical Engineering, Keio University, Yokohama, 223-8522, Japan
}%


\date{\today}

\begin{abstract}
  Recently, anomalous subdiffusion, aging, and scatter of the diffusion
  coefficient have been reported in many single-particle-tracking experiments,
  though origins of these behaviors are still elusive. Here, as a model to
  describe such phenomena, we investigate a Langevin equation with diffusivity
  fluctuating between a fast and a slow state. We assume that the sojourn time
  distributions of these two states are given by power laws. It is shown that,
  for a non-equilibrium ensemble, the ensemble-averaged mean square displacement
  (MSD) shows transient subdiffusion. In contrast, the time-averaged MSD shows
  normal diffusion, but an effective diffusion coefficient transiently shows
  aging behavior. The propagator is non-Gaussian for short time, and converges
  to a Gaussian distribution in a long time limit; this convergence to Gaussian
  is extremely slow for some parameter values. For equilibrium ensembles, both
  ensemble-averaged and time-averaged MSDs show only normal diffusion, and thus
  we cannot detect any traces of the fluctuating diffusivity with these
  MSDs. Therefore, as an alternative approach to characterize the fluctuating
  diffusivity, the relative standard deviation (RSD) of the time-averaged MSD is
  utilized, and it is shown that the RSD exhibits slow relaxation as a signature
  of the long-time correlation in the fluctuating diffusivity.
  Furthermore, it is shown that the RSD is related to a non-Gaussian parameter
  of the propagator. To obtain these theoretical results, we develop a two-state
  renewal theory as an analytical tool.
\end{abstract}

\maketitle


\section {Introduction}

Temporal and/or spacial heterogeneities of diffusivity have been reported in
various systems such as entangled polymer systems \cite{doi78, uneyama15},
macromolecular diffusion in cells \cite{parry14, leith12},
and supercooled liquids \cite{yamamoto98a, yamamoto98b}.
In \cite{doi78, uneyama15}, it is shown that the center-of-mass motion of
reptation dynamics in entangled polymer systems exhibits temporal fluctuations
of diffusivity.
In \cite{parry14}, it is reported that particles in a bacterial cytoplasm show a
fast and a slow diffusion mode.
%
%
Similarly, one-dimensional diffusion of a eukaryotic transcription factor
along DNA strands has a fast and a slow mode with a local variation of the
diffusion coefficient (DC) along the DNA \cite{leith12}.
%
%
%
Molecular dynamics simulations of supercooled liquids reported in
\cite{yamamoto98a, yamamoto98b} show a spatial heterogeneity with roughly two
values of diffusivity. As a result of such a spacial heterogeneity of
diffusivity, a tagged particle will display temporally fluctuating diffusivity.
Also, fluctuating diffusivity would be observed in systems with dynamical
heterogeneity (i.e., coexistence of spatial and temporal heterogeneities) such
as various glass formers \cite{berthier11}.
%



Although, as a phenomenological model of single-particle diffusion in spatial
and temporal heterogeneities, trap models---i.e., continuous-time random walks
(CTRWs) \cite{he08, neusius09, weigel11, meroz10, miyaguchi11c, miyaguchi13,
  jeon13,schulz14} and quenched trap models \cite{bouchaud90
}---have been widely used, there have been increasing reports in which trap
models fail to describe complex systems with heterogeneities.  For example, in
\cite{manzo15}, Manzo et al. show that receptor diffusion on cell membranes is
not consistent with the CTRW, and proposed the annealed transit time model
(ATTM) which is a Brownian motion with fluctuating diffusivity. In the ATTM, the
fluctuating diffusivity is considered to be originated from a local variation of
the DC \cite{massignan14, manzo15}, which is observed in many biological
experiments \cite{english11, kuhn11, cutler13, masson14}. Moreover, a single
polymer in an entangled polymer solution is transiently trapped in a virtual
tube comprised by surrounding polymers \cite{doi78, doi86}. Therefore, it is
natural to expect that this kind of trapped motion could be described by a CTRW,
but this is not the case; instead, the center of mass motion of a single
entangled polymer is clearly described by a Langevin equation with fluctuating
diffusivity (LEFD) \cite{uneyama15}. An elaborate simulation study on
supercooled liquids in \cite{helfferich14a, helfferich14b} also shows that a
tagged particle trajectory slightly deviates from that of CTRW.




In financial mathematics, stochastic differential equations called stochastic
volatility models, in which the DC follows other stochastic differential
equations, have been extensively studied \cite{stein91, heston93, scott97}.
Moreover, the underdamped version of the LEFD is analyzed in \cite{rozenfeld98,
  luczka00}, and a special type of the overdamped LEFD, a random walk with
diffusing diffusivity (i.e., the DC follows a diffusion equation), is studied
recently in \cite{mykyta14}. The ATTM stated above is also a special class of
the overdamped LEFD, in which the DC is distributed according to a power
law. Along with these models, a two-state LEFD, in which diffusivity fluctuates
between two values, $D_+$ and $D_-$, should be important. This is because, in
some experiments including those stated above \cite{yamamoto98a, yamamoto98b,
  knight09, leith12, parry14}, it is reported that the DC has roughly two
distinct values. Such a two-state LEFD as well as a general form of the
overdamped LEFD are studied for the case of equilibrium processes in
\cite{uneyama15}.  Here, we study the overdamped LEFD with dichotomous DCs for
both equilibrium and non-equilibrium processes.
Furthermore, the LEFD is considered to be important for intermittent search
\cite{loverdo09,benichou11}. In fact, LEFD-like systems with dichotomous DCs are
studied as an efficient search strategy for finding a target
\cite{reingruber09,reingruber10}. In these studies, however, Markovian
switchings between the DCs are used, whereas we investigate general
non-Markovian switchings.




A time-averaged mean square displacement (MSD) is frequently used in experiments
\cite{golding06,burov11, jeon11, he08} and utilized even in some molecular
dynamics simulations \cite{akimoto11, yamamoto14, yamamoto15}. However, the
time-averaged MSD (TMSD) of the LEFD shows only normal diffusion, and thus it is
impossible to detect and characterize the fluctuating diffusivity. Besides, the
instantaneous DC, $D(t)$, is quite difficult to measure accurately in
experiments, since $D(t)$ is multiplied by the thermal noise in the Langevin
equation [see Eq.~(\ref{e.lefd})]. Then, how can we characterize $D(t)$ from
experimental data? In this article, we show that the relative standard deviation
(RSD) of the TMSD is a useful observable to elucidate the fluctuating
diffusivity, because the RSD is closely related to the auto-correlation function
(ACF), $\left\langle D(t)D(t') \right\rangle$. An important point is that the
RSD can be calculated from dozens of trajectories $\bm{r}(t)$, and it is not
necessary to directly measure $D(t)$.

It is also shown that, in a certain limit ($D_- \to 0$), the LEFD shares many
properties with the CTRW: subdiffusion in the ensemble-averaged MSD (EMSD),
aging in the TMSD, and non-Gaussianity. These properties are common in
single-particle-tracking experiments \cite{golding06,burov11,jeon11} and
molecular dynamics simulations \cite{akimoto11, yamamoto14, yamamoto15}, and
have been a basis of the CTRW modeling of such systems \cite{he08, neusius09,
  weigel11, meroz10, miyaguchi11c, miyaguchi13, jeon13,schulz14}. Therefore, the
LEFD is another candidate for modeling such anomalous systems. In this article,
various anomalous features mentioned above are checked by numerical simulations
for the one-dimensional LEFD (the theoretical results are applicable to
$n$-dimensional systems).

Furthermore, we develop a generalized renewal theory as an analytical
tool. Systems studied in the usual renewal theory have only a single state
\cite{godrche01}, while the LEFD in this article has two states. Therefore, it is
necessary to generalize the usual (single-state) renewal theory to the one with
two states. Some previous works have already worked on such a generalization
\cite{cox62, goychuk03, akimoto15, uneyama15}, but, in this article, a much
broader framework of the two-state renewal theory, which is also called an
alternating renewal theory in \cite{cox62}, is presented. For example, the
fractions of the states, transition probabilities, and probability density
functions (PDFs) of a forward recurrence time are investigated.

This paper is organized as follows. In Sec.~\ref{s.model}, we define the LEFD,
switching rules between the two states (i.e., sojourn time PDFs of these
states), and equilibrium as well as non-equilibrium ensembles. In
Sec.~\ref{s.general_theory}, analytical results for arbitrary sojourn time PDFs
are described, whereas, in
Secs.~\ref{s.case_study_fraction}--~\ref{s.case_study_rsd}, case studies for
power-law sojourn time PDFs are presented, and relations between the LEFD and
the CTRW are also pointed out here. Finally, Sec.~\ref{s.discussion} is devoted
to a discussion. In the Appendices, we summarize some technical matters,
including simulation details and a relation between the RSD and a non-Gaussian
parameter.

\section {Langevin equation with fluctuating Diffusivity}\label{s.model}

In this section, the LEFD with a fast and a slow diffusion state is
defined. Then, we adopt non-Markovian switching rules between the two states
through sojourn time PDFs of these states. Also, equilibrium and non-equilibrium
initial ensembles are defined with first sojourn time PDFs.

\subsection {Definition of the LEFD}

The LEFD is given by the following equation:
\begin{equation}
  \label{e.lefd}
  \frac {d\bm{r}(t)}{dt} = \sqrt{2 D(t)}\, \bm{\xi}(t),
\end{equation}
where $\bm{r}(t)$ is the $n$-dimensional position of the diffusing particle, and
$\bm{\xi}(t)$ is Gaussian white noise vector with
\begin{equation}
  \left\langle \bm{\xi}(t) \right\rangle = \bm{0}, \qquad
  \left\langle \xi_i(t)\xi_j(t') \right\rangle = \delta_{ij}\delta(t-t').
\end{equation}
In contrast, the time-dependent DC, $D(t)$, is allowed to be a non-Markovian
stochastic process, whereas we assume that $D(t)$ and $\bm{\xi}(t)$ are
statistically independent. Equation (\ref{e.lefd}) is a general definition of an
isotropic LEFD. An anisotropic version, which can be utilized to describe the
center-of-mass motion of an entangled polymer in a reptation model, is analyzed
in \cite{uneyama15}.


\subsection {Two-state system}\label{s.two-state-system}

\begin{figure}[]
  \centerline{\includegraphics[width=7.8cm]{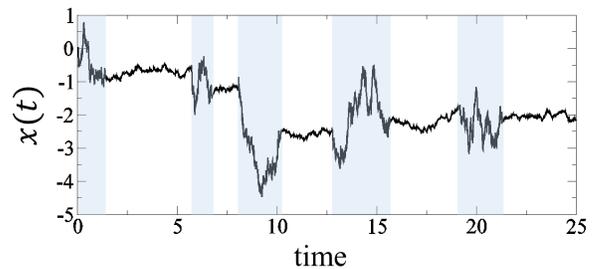}}
  \caption{\label{f.trajectory}An example of trajectory $x(t)$ of the two-state
    LEFD. The regions with shaded and white backgrounds correspond to the fast
    and the slow modes, respectively. The ratio $D_+/D_-$ is set as $D_+/D_- =
    50$.}
\end{figure}


In this article, we consider a two-state process, i.e., the state of the system
is alternating between two modes, labeled $+$ and $-$. More precisely, it is
assumed that the DC is given by $D(t)=D_+$ if the state is $+$ at time $t$, and
$D(t)=D_-$ if the state is $-$ at time $t$. We also assume that $D_- \leq D_+$
for clarity, though this restriction is not necessary for the theoretical
analysis.  A trajectory of this two-state LEFD is displayed in
Fig.~\ref{f.trajectory}.


A switching rule between the two states can be given by sojourn time PDFs of
these states. Namely, the sojourn times $\tau$ for $+$ and $-$ states are random
variables following different PDFs, $\rho_+(\tau)$ and $\rho_-(\tau)$. We assume
that these PDFs follow power-law distributions (though the results in
Sec.~\ref{s.general_theory} are applicable for arbitrary PDFs):
\begin{equation}
  \label{e.rho(t)}
  \rho_\pm (\tau)
  \underset{\tau\to\infty}{\simeq} 
  \frac {a_{\pm}}{|\Gamma (-\alpha_{\pm})|\tau^{1+\alpha_\pm}},
\end{equation}
where $a_{\pm}$ is a scale factor and $\Gamma (-\alpha_{\pm})$ is the gamma
function.
Asymptotic forms of the Laplace transforms $\hat{\rho}_{\pm} (s) :=
\int_0^{\infty} d\tau\, e^{-\tau s} \rho_{\pm}(\tau)$ at small $s$ are given by
\begin{alignat}{2}
  \label{e.rho(s).asymptotic.alpha<1}
  \hat{\rho}_{\pm}(s) & \underset{s \to 0}{=} 1 - a_{\pm} s^{\alpha_{\pm}} + O(s)
  \quad               & \alpha_{\pm} \in (0,1), \\[0.0cm]
  \label{e.rho(s).asymptotic.alpha>1}
  \hat{\rho}_{\pm}(s) & \underset{s \to 0}{=}  1
  - \mu_{\pm} s + a_{\pm} s^{\alpha_{\pm}}  + O(s^2)
  \quad               & \alpha_{\pm} \in (1,2),
\end{alignat}
where $\mu_\pm$ is the mean sojourn time of the state $\pm$ and $O(s^n)$ is
Landau's notation.  Note that $\mu_{\pm}$ does not exist for $\alpha_{\pm} \in
(0,1)$. Moreover, in Eqs.~(\ref{e.rho(s).asymptotic.alpha<1}) and
(\ref{e.rho(s).asymptotic.alpha>1}), we assume that the sub-leading corrections
to Eq.~(\ref{e.rho(t)}) decay faster than $\tau^{-2}$ and $\tau^{-3}$,
respectively \cite{bardou01}; these assumptions are critical in the following
analysis \footnote{For the case(1), for example, we frequently use the expansion
  $1- \hat{\rho}(s) = 1- \hat{\rho}_+(s)\hat{\rho}_-(s) \simeq a_+s^{\alpha_+} +
  a_-s^{\alpha_-} + O(s, s^{\alpha_+ + \alpha_-})$. This equation is not
  necessarily valid if there are second sublinear modes: $\hat{\rho}_{\pm}(s)
  \simeq 1 - a_{\pm} s^{\alpha_{\pm}} - a_{\pm}' s^{\alpha'_{\pm}} + O(s)$},
although some important PDFs such as the L\'evy's stable law with $\alpha<1/2$
do not satisfy this assumption.


Here, let us comment on a possible origin of the power-law sojourn time PDFs. It
is well known that, for crowded systems such as supercooled liquids, clusters of
fast (mobile) and slow (immobile) particles are formed \cite{ediger00,
  weeks00}. For example, a slow particle diffuses in a slow cluster, eventually
traverses a boundary between clusters, and goes into a fast cluster. Thus, the
sojourn time PDF of the slow state may be given by a PDF of the first passage
time to the boundary. Such a first passage time PDF in a compact domain is
usually given by a power law with an exponential cutoff \cite{redner01}. This
exponential cutoff results in a crossover from anomalous to normal behaviors,
but the crossover time could be very long \cite{mantegna94, miyaguchi11c,
  miyaguchi13}; thus, the pure power law might be a reasonable assumption
(though, this reasoning is still too simplistic, since the clusters themselves
are changing with time).


Here, we consider the following six cases for the values of the exponents
$\alpha_+$ and $\alpha_-$:
\begin{itemize}
  \setlength{\leftskip}{2.5cm}
  \item [case (1--1)]: $\alpha_{\pm} \in (0,1)$ with $\alpha_+ < \alpha_-$
  \\[-0.52cm]
  \item [case (1--2)]: $\alpha_{\pm} \in (0,1)$ with $\alpha_+ > \alpha_-$
  \\[-0.52cm]
  \item [case (2--1)]: $\alpha_{\pm} \in (1,2)$ with $\alpha_+ < \alpha_-$
  \\[-0.52cm]
  \item [case (2--2)]: $\alpha_{\pm} \in (1,2)$ with $\alpha_+ > \alpha_-$
  \\[-0.52cm]
  \item [case (3--1)]: $\alpha_{+} \in (0,1)$ and $\alpha_{-} \in (1,2)$
  \\[-0.52cm]
  \item [case (3--2)]: $\alpha_{+} \in (1,2)$ and $\alpha_{-} \in (0,1)$
\end{itemize}
But, essentially, there are only three cases, because, for example, the results
for the case (1--2) can be obtained from those of the case (1--1) by
substituting the $\pm$ signs in subscripts for $\mp$ signs. Nevertheless, there
are qualitative differences between the cases (1--1) and (1--2), and therefore
we use the classification into six cases for clarity.  We also use a generic
term ``the case (1)'' to refer to both cases (1--1) and (1--2). The same
notations such as the ``case (2)'' and ``case (3)'' are also used.


\subsection {Initial ensembles}

Initial ensembles are specified by first sojourn time PDFs. We assume that the
process starts at $t=0$, and then, at some time $t= \tau$, the first transition
from one state to the other occurs. Let us write the PDF of this first sojourn
time, $\tau$, as $w_{+}^{0}(\tau) + w_{-}^{0}(\tau)$. Here, $w_{\pm}^{0}(\tau)$
is defined as $w_{\pm}^{0}(\tau) := p_{\pm}^0\rho_{\pm}^{0}(\tau)$, where
$p_{\pm}^0$ is an initial fraction and $\rho_{\pm}^{0}(\tau)$ is a first sojourn
time PDF given that the initial state is $\pm$.  We also use a vector notation
$\bm{w}^0(\tau) := (w_+^{0}(\tau),w_-^{0}(\tau))$, with which we can completely
specify the initial ensembles.

For the equilibrium initial ensemble, we set $\rho_{\pm}^{0}(\tau) =
\rho_{\pm}^{\mathrm{eq}}(\tau)$, which is given by its Laplace transform
\cite{godrche01} (see also Appendix \ref{s.pdf-elapsed-time}):
\begin{equation}
  \label{e.init-deinsity-equilibrium}
  \hat{\rho}_{\pm}^{\mathrm{eq}}(s) = \frac {1 - \hat{\rho}_{\pm}(s)}{\mu_{\pm}s}.
\end{equation}
Hence, this PDF exists only if $\mu_{\pm}$ is finite.
On the other hand, we set $\rho_{\pm}^{0}(\tau) = \rho_{\pm}(\tau)$ for a
non-equilibrium initial ensemble, i.e., the first sojourn time PDF,
$\rho_{\pm}^{0}(\tau)$, is the same as the PDF of the second and later sojourn
times $\rho_{\pm}(\tau)$ (this is a typical non-equilibrium initial ensemble
used in the renewal theory and the CTRW theory).

To define the initial ensembles, we also have to specify initial fractions of
the two states, $p_{\pm}^0$. In the equilibrium initial ensembles, these
fractions are given by (see Appendix \ref{s.pdf-elapsed-time})
\begin{equation}
  \label{e.eq-ensemble-1}
  p_{\pm}^0=
  \frac {\mu_{\pm}}\mu
  =: p_{\pm}^{\mathrm{eq}}, 
\end{equation}
where $\mu$ is defined by $\mu:= \mu_+ + \mu_-$.  For the non-equilibrium
initial ensembles, we leave $p_{\pm}^0\in [0,1]$ arbitrary, because the
asymptotic behavior of the system is independent of $p_{\pm}^0$ as shown in the
following sections.

Thus, the equilibrium ensemble, $\hat{w}_{\pm}^{\mathrm{eq}}(s)$, is
given by
\begin{equation}
  \label{e.eq-ensemble-2}
  \hat{w}_{\pm}^{\mathrm{eq}}(s)=
  p_{\pm}^{\mathrm{eq}}\hat{\rho}_{\pm}^{\mathrm{eq}}(s)
  =
  p_{\pm}^{\mathrm{eq}}\frac {1 - \hat{\rho}_{\pm}(s)}{\mu_{\pm} s}.
\end{equation}
We denote the ensemble average in terms of this equilibrium ensemble as
$\left\langle \cdot \right\rangle_{\mathrm{eq}}$. Similarly, the following
non-equilibrium ensembles shall be considered:
\begin{equation}
  \hat{w}_{\pm}^{\mathrm{neq}}(s)
  =
  p_{\pm}^0\hat{\rho}_{\pm}(s).
  \label{noneq_initial}
\end{equation}
We depict the ensemble average in terms of this non-equilibrium ensemble
as $\left\langle \cdot \right\rangle_{\mathrm{neq}}$. Note that, in this
notation $\left\langle \cdot \right\rangle_{\mathrm{neq}}$, we do not explicitly
show the dependence on the initial fractions $p_{\pm}^0$.
In numerical simulations, we use two non-equilibrium ensembles: one that starts
from $+$ state [i.e., $(p_+^0, p_-^0) = (1,0)$] and the other that starts from
$-$ state [i.e., $(p_+^0, p_-^0) = (0,1)$].
When we use the bracket without a subscript $\left\langle \cdot \right\rangle$,
it means that the average is taken over an arbitrary initial ensemble
$w^0_{\pm}(\tau)$.


\section {General theory}\label{s.general_theory}

In this section, we derive general formulas for equilibrium and non-equilibrium
LEFD with arbitrary sojourn time PDFs, $\rho_{\pm}(\tau)$. In particular, we
study fractions of the states, the EMSD, the ensemble-averaged TMSD (ETMSD), the
propagator, and the RSD of the TMSD.

\subsection {Transition probabilities}

We start with transition probabilities $W_{hh'} (t|\bm{w}^0)$ where $h$ and $h'$
stand for the states: $h, h' = \pm$. More precisely, $W_{hh'}(t|\bm{w}^0)$ is a
conditional joint probability that the state is $h$ at time $t=0$ and the state
is $h'$ at time $t$, given that the process starts with the initial ensemble
$\bm{w}^0(\tau)$. Let us rewrite $W_{hh'} (t|\bm{w}^0)$ as
\begin{align}
  \begin{split}
    W_{\pm\pm} (t|\bm{w}^0) &=\sum_{n=0}^{\infty} Q_{\pm,2n}(t|\bm{w}^0), \\
    W_{\pm\mp} (t|\bm{w}^0) &=\sum_{n=1}^{\infty} Q_{\pm,2n-1}(t|\bm{w}^0),
  \end{split}
\end{align}
where $Q_{\pm,n}(t|\bm{w}^0)$ is a joint probability that the state is $\pm$ at
time $t=0$ and there is $n$ transitions until time $t$, given that the process
starts with $\bm{w}^0(\tau)$. The Laplace transforms are given by
\begin{align}
  \label{e.W(s)}
  \begin{split}
    \hat{W}_{\pm\pm} (s|\bm{w}^0) & = \sum_{n=0}^{\infty} \hat{Q}_{\pm,2n}(s|\bm{w}^0), \\
    \hat{W}_{\pm\mp} (s|\bm{w}^0) & =\sum_{n=1}^{\infty} \hat{Q}_{\pm,2n-1}(s|\bm{w}^0).
  \end{split}
\end{align}
The probability $Q_{\pm,n}(t|\bm{w}^0)$ is given as follows:
\begin{align}
  \begin{split}
    Q_{\pm,0} (t|\bm{w}^0)
    &= \mathcal{I}w_{\pm}^{0} (t),
    \\
    \label{e.Q(t)}
    Q_{\pm,2n}(t|\bm{w}^0)
    &= w_{\pm}^{0} * \rho^{*(n-1)} * \rho_{\mp} * \mathcal{I}\rho_{\pm}(t),
    \\
    Q_{\pm,2n-1}(t|\bm{w}^0)
    &= w_{\pm}^{0} * \rho^{*(n-1)} * \mathcal{I}\rho_{\mp} (t), 
  \end{split}
\end{align}
where $n=1,2,\dots$, and an operator $\mathcal{I}$ is defined as
$\mathcal{I}f(t) := \int_t^{\infty} dt' f(t')$.  Moreover, we define a PDF
$\rho(t)$ by a convolution as
\begin{equation}
  \rho(t) := \rho_+ * \rho_- (t) = \int_0^t d\tau \rho_+ (t-\tau)\rho_-(\tau),
\end{equation}
and also $n$-time convolution $\rho^{*n}(t)$ as
\begin{equation}
  \rho^{*n}(t) := \underbrace{\rho*\dots*\rho}_{n \text{times}}(t).  
\end{equation}

Then, the Laplace transforms of Eqs.~(\ref{e.Q(t)}) are given by
\begin{align}
  \label{e.Q(s)}
  \begin{split}
    \hat{Q}_{\pm,0} (s|\bm{w}^0)
    &= \frac {p_{\pm}^0-\hat w^0_{\pm} (s)}{s},
    \\[0.0cm]
    \hat{Q}_{\pm,2n}(s|\bm{w}^0)
    &= \hat w_{\pm}^0(s) \hat{\rho}^{n-1}(s)\hat{\rho}_{\mp}(s) \frac{1-\hat{\rho}_{\pm}(s)}{s},
    \\[0.0cm]
    \hat{Q}_{\pm,2n-1}(s|\bm{w}^0)
    &= \hat{w}_{\pm}^0(s) \hat{\rho}^{n-1}(s)\frac {1-\hat{\rho}_{\mp}(s)}{s},
  \end{split}
\end{align}
where $n=1,2,\dots$ (A different derivation is given in Appendix
\ref{s.app.Q}). Putting Eq.~(\ref{e.Q(s)}) into Eq.~(\ref{e.W(s)}), we have
\begin{align}
  \label{e.W(s).2}
  \begin{split}
    \hat{W}_{\pm\pm} (s|\bm{w}^0) &= 
    \frac {p_{\pm}^0}{s} - \frac {\hat{w}_{\pm}^0(s) }{s}
    \frac{1-\hat{\rho}_{\mp}(s)}{1- \hat{\rho}(s)},
    \\[.0cm]
    \hat{W}_{\pm\mp} (s|\bm{w}^0) &= 
    \frac {\hat{w}_{\pm}^0(s) }{s}
    \frac{1-\hat{\rho}_{\mp}(s)}{1- \hat{\rho}(s)}.
  \end{split}
\end{align}
A special case [$p_+^0 = 1$, and $\hat{w}_+^0(s) =
\hat{\rho}_{+}^{\mathrm{eq}}(s)$] of these equations is studied in
\cite{cox62}. Note that $\hat{W}_{\pm\pm} (s|\bm{w}^0) + \hat{W}_{\pm\mp}
(s|\bm{w}^0) = p_{\pm}^0/s$, which means the normalization: $W_{\pm\pm}
(t|\bm{w}^0) + W_{\pm\mp} (t|\bm{w}^0) = p_{\pm}^0$.

Inserting Eqs.~(\ref{e.eq-ensemble-1}) and (\ref{e.eq-ensemble-2}) into
Eq.~(\ref{e.W(s).2}), we have the transition probabilities for the equilibrium
processes
\begin{align}
  \label{e.W(s).eq}
  \begin{split}
    \hat{W}_{\pm\pm}^{\mathrm{eq}} (s)
    &:= \hat{W}_{\pm\pm} (s|\bm{w}^{\mathrm{eq}})\\[0.0cm]
    &=
    \frac {p_{\pm}^{\mathrm{eq}}}{s} - \frac 1{\mu s^2}
    \frac{\left[1-\hat{\rho}_+(s)\right]\left[1-\hat{\rho}_-(s)\right]}
    {1- \hat{\rho}(s)},
    \\[.0cm]
    \hat{W}_{\pm\mp}^{\mathrm{eq}} (s)
    &:= \hat{W}_{\pm\mp} (s|\bm{w}^{\mathrm{eq}})\\[0.0cm]
    &=
    \frac 1{\mu s^2}
    \frac{\left[1-\hat{\rho}_+(s)\right]\left[1-\hat{\rho}_-(s)\right]}
    {1- \hat{\rho}(s)}.
  \end{split}
\end{align}
From the second equation, we have $\hat{W}_{+-}^{\mathrm{eq}} (s) =
\hat{W}_{-+}^{\mathrm{eq}} (s)$, i.e., the detailed balance holds. The above
equations (\ref{e.W(s).eq}) are already reported in \cite{goychuk03}.

On the other hand, the transition probabilities for the non-equilibrium
processes are given by
\begin{align}
  \label{e.W(s).neq}
  \begin{split}
    \hat{W}_{\pm\pm}^{\mathrm{neq}} (s)
    &:=
    \hat{W}_{\pm\pm}(s|\bm{w}^{\mathrm{neq}}) 
    =
    \frac {p_{\pm}^0}{s}
    \frac{1-\hat{\rho}_{\pm}(s)}{1- \hat{\rho}(s)},
    \\[.0cm]
    \hat{W}_{\pm\mp}^{\mathrm{neq}} (s)
    &:=
    \hat{W}_{\pm\mp} (s|\bm{w}^{\mathrm{neq}})
    =
    \frac {p_{\pm}^0}{s}
    \frac{1-\hat{\rho}_{\mp}(s)}{1- \hat{\rho}(s)} \hat{\rho}_{\pm}(s),
  \end{split}
\end{align}
where we inserted Eq.~(\ref{noneq_initial}) into Eq.~(\ref{e.W(s).2}). From the
second equation, we see that $\hat{W}_{+-}^{\mathrm{neq}} (s) \not\equiv
\hat{W}_{-+}^{\mathrm{neq}} (s)$, i.e., the detailed balance breaks down.

\subsection {Fraction of the state}\label{s.general.fraction}

Ensemble averages of some single-time observables can be obtained from the
fractions of the states [e.g., see Eqs.~(\ref{e.D(t).general}) and
(\ref{e.EMSD(s).general})].  Here, we define the fraction of the state,
$p_{\pm}(t|\bm{w}^0)$, as the probability being in the state $\pm$ at time $t$,
given that the initial ensemble is $w^0_{\pm}(\tau)$. We have
$p_{\pm}(0|\bm{w}^0) = p_{\pm}^0$ and
\begin{equation}
  \label{e.fraction.p(t)}
  p_{\pm} (t|\bm{w}^0) = W_{\pm\pm} (t|\bm{w}^0) + W_{\mp\pm}(t|\bm{w}^0).
\end{equation}
Therefore, from Eqs.~(\ref{e.W(s).2}) and (\ref{e.fraction.p(t)}), we have the
following general expression for the Laplace transform of the fraction:
\begin{equation}
  \label{e.fraction.p(s)}
  \hat{p}_{\pm}(s|\bm{w}^0) = \frac {p_{\pm}^0}{s}
  - \frac {\hat{w}_{\pm}^0(s)}{s} \frac {{1-\hat{\rho}_{\mp}(s)}}{1-\hat{\rho}(s)}
  + \frac {\hat{w}_{\mp}^0(s)}{s} \frac {{1-\hat{\rho}_{\pm}(s)}}{1-\hat{\rho}(s)}.
\end{equation}
If there exist mean sojourn times $\mu_{\pm}$, we have from
Eq.~(\ref{e.fraction.p(s)})
\begin{equation}
  \label{e.p(t)->p^eq(t)}
  \lim_{s \to 0} s p_{\pm}(s|\bm{w}^0)
  =
  p_{\pm}^{\mathrm{eq}}.
\end{equation}
Thus, the fractions $p_{\pm}(t|\bm{w}^0)$ tend to the equilibrium fractions
$p_{\pm}^{\mathrm{eq}}$ as $t\to \infty$. On the other hand, if a mean sojourn
time, $\mu_{+}$ or $\mu_{-}$, does not exist, the system never reaches an
equilibrium; that is, there is no equilibrium state.


Using Eqs.~(\ref{e.W(s).eq}) and (\ref{e.fraction.p(t)}), we have the
fractions for the equilibrium processes as
\begin{align}
  \label{e.fraction.p(s).eq}
  \hat{p}_{\pm}^{\mathrm{eq}}(s)
  :=
  \hat{p}_{\pm}(s|\bm{w}^{\mathrm{eq}})
  &=
  \hat{W}_{\pm\pm}^{\mathrm{eq}} (s) + \hat{W}_{\mp\pm}^{\mathrm{eq}}(s)
  =
  \frac {p_{\pm}^{\mathrm{eq}}}{s},
\end{align}
and hence the equilibrium fractions are stationary: $p_{\pm}^{\mathrm{eq}}(t)
\equiv p_{\pm}^{\mathrm{eq}}$.
%
On the other hand, from Eqs.~(\ref{e.W(s).neq}) and (\ref{e.fraction.p(t)}), the
non-equilibrium fractions, $\hat{p}_{\pm}^{\mathrm{neq}}(s)$, are given by
\begin{align}
  \label{e.fraction.p(s).noneq}
  \hat{p}_{\pm}^{\mathrm{neq}} (s)
  :=
  \hat{p}_{\pm}(s|\bm{w}^{\mathrm{neq}})
  &=
  \hat{W}_{\pm\pm}^{\mathrm{neq}} (s) + \hat{W}_{\mp\pm}^{\mathrm{neq}}(s)
  \notag\\[0.0cm]
  &=
  \frac {p_{\pm}^0 + p_{\mp}^0\hat{\rho}_{\mp}(s)}{s}
  \frac {1 - \hat{\rho}_{\pm}(s)}{1-\hat{\rho}(s)}.
\end{align}


\subsection {Ensemble average of the MSD}\label{s.emsd.general}

Using Eq.~(\ref{e.lefd}) and $\left\langle \bm{\xi}(t')\cdot\bm{\xi}(t'')
\right\rangle = n \delta(t'-t'')$, we have the following expression for the
EMSD $\left\langle \delta\bm{r}^2(t) \right\rangle$:
\begin{equation}
  \label{e.emsd}
  \left\langle \delta\bm{r}^2(t) \right\rangle
  =
  2n \int_{0}^{t}dt' \left\langle D(t') \right\rangle,
\end{equation}
where $\delta\bm{r}(t)$ is a displacement $\delta\bm{r}(t) := \bm{r}(t) -
\bm{r}(0)$.  Hence, the EMSD can be obtained by integrating $\left\langle D(t)
\right\rangle$, which is in turn given by
\begin{align}
  \left\langle D(t) \right\rangle
  &=
  D_+ p_+(t|\bm{w}^0) + D_- p_-(t|\bm{w}^0)\notag\\[0.0cm]
  \label{e.D(t).general}
  &=
  D_- + (D_+ - D_-) p_+(t|\bm{w}^0).
\end{align}
Then, the Laplace transform  is given by
\begin{equation}
  \label{e.D(s).general}
  \langle \hat{D}(s) \rangle
  =
  \frac {D_-}{s} + (D_+ - D_-) \hat{p}_+(s|\bm{w}^0),
\end{equation}
and therefore, from Eqs.~(\ref{e.emsd}) and (\ref{e.D(s).general}), the Laplace
transform of the EMSD is given by
\begin{equation}
  \mathcal{L}
  \left[\left\langle \delta\bm{r}^2(t)
  \right\rangle\right](s)
  =
  \label{e.EMSD(s).general}
  2n\left[
  \frac {D_-}{s^2} +
  (D_+ - D_-) \frac {\hat{p}_+(s|\bm{w}^0)}{s} 
  \right].
\end{equation}
Inserting Eq.~(\ref{e.fraction.p(s)}) into Eq.~(\ref{e.EMSD(s).general}), we
obtain a general expression of the EMSD in terms of the sojourn time PDFs.


For the equilibrium ensemble, we have from Eq.~(\ref{e.D(t).general})
\begin{equation}
  \label{e.<D(s)>_eq}
  \langle D(t) \rangle_{\mathrm{eq}}
  =
  D_{+} p_+^{\mathrm{eq}} + D_{-} p_-^{\mathrm{eq}} =: D_{\mathrm{eq}}.
\end{equation}
Namely, the equilibrium processes are stationary as expected. From
Eq.~(\ref{e.emsd}), we have
\begin{equation}
  \label{e.emsd(t).eq}
  \left\langle
  \delta\bm{r}^2(t) 
  \right\rangle_{\mathrm{eq}}
  =
  2n D_{\mathrm{eq}} t,  
\end{equation}
and hence any equilibrium processes show normal diffusion (this result is
consistent with the same property obtained for the underdamped LEFD
\cite{rozenfeld98,luczka00}). That is, the second moment of the particle
position $\bm{r}(t)$ cannot detect any anomaly, i.e., fluctuating diffusivity,
of the equilibrium systems, and therefore it is necessary to study some higher
order moments instead.


For the non-equilibrium ensemble, using Eqs.~(\ref{e.fraction.p(s).noneq}) and
Eq.~(\ref{e.D(t).general}), we have
\begin{equation}
  \label{e.D(t).general.neq}
  \langle D(t) \rangle_{\mathrm{neq}}
  =
  D_- + (D_+ - D_-) p_+^{\mathrm{neq}}(t).
\end{equation}
Moreover, by using Eqs.~(\ref{e.fraction.p(s).noneq}) and
(\ref{e.EMSD(s).general}), the Laplace transform of the non-equilibrium EMSD is
given by
\begin{equation}
  \mathcal{L}
  \left[\left\langle \delta\bm{r}^2(t)
  \right\rangle_{\mathrm{neq}}\right]\!(s)
  =
  \label{e.EMSD.general.noneq}
  2n\left[
  \frac {D_-}{s^2} +
  (D_+ - D_-) \frac {\hat{p}_+^{\mathrm{neq}}(s)}{s} 
  \right]\!.
\end{equation}


\subsection {Ensemble average of the TMSD}

The TMSD, which is frequently used in single particle tracking experiments
\cite{golding06,burov11,jeon11, he08} as well as in molecular dynamics
simulations \cite{akimoto11, yamamoto14, yamamoto15}, is defined as
\begin{equation}
  \label{e.def.tmsd}
  \overline{\delta\bm{r}^2}(\Delta; t)
  :=
  \frac {1}{t-\Delta} \int_{0}^{t-\Delta} dt'
  \left[ \bm{r}(t'+\Delta) - \bm{r}(t') \right]^2.
\end{equation}
Here, $t$ is the total measurement time, and $\Delta$ is the lag
time. Interestingly, a quantity similar to the TMSD is used as an order
parameter for glassy systems \cite{hedges09}. As with the derivation of
Eq.~(\ref{e.emsd}), Eq.~(\ref{e.def.tmsd}) can be rewritten as
\begin{align}
  \left\langle \overline{\delta\bm{r}^2}(\Delta; t) \right\rangle
  &=
  \frac {1}{t-\Delta} \int_{0}^{t-\Delta} dt'
  \left\langle \left[ \bm{r}(t'+\Delta) - \bm{r}(t') \right]^2 \right\rangle
  \notag\\[0.0cm]
  \label{e.etmsd(t).1}
  &=
  \frac {2n}{t-\Delta} \int_{0}^{t-\Delta} dt'
  \int_{t'}^{t'+\Delta}d\tau  \left\langle D(\tau) \right\rangle.
\end{align}
For the equilibrium ensembles, using Eqs.~(\ref{e.<D(s)>_eq}) and
(\ref{e.etmsd(t).1}), we have
\begin{equation}
  \label{e.etmsd(t).eq}
  \left\langle \overline{\delta \bm{r}^2}(\Delta; t) \right\rangle_{\mathrm{eq}}
  =
  2n D_{\mathrm{eq}} \Delta,
\end{equation}
and hence the ETMSD is equivalent to the EMSD [Eq.~(\ref{e.emsd(t).eq})].


To deal with general non-equilibrium processes, we rewrite
Eq.~(\ref{e.etmsd(t).1}) by using Eq.~(\ref{e.emsd}) as
\begin{equation}
  \label{e.etmsd(t).2}
  \left\langle \overline{\delta\bm{r}^2}(\Delta; t) \right\rangle
  \underset{\Delta \ll t}{\simeq}
  \frac {1}{t} \int_0^t dt'
  \Bigl[
  \left\langle \delta\bm{r}^2(t'+\Delta) \right\rangle
  -
  \left\langle \delta\bm{r}^2(t') \right\rangle
  \Bigr]
\end{equation}
where we assume that $\Delta \ll t$. The Laplace transform (with respect to $t$)
of the integral in Eq.~(\ref{e.etmsd(t).2}) gives
\begin{align}
  \mathcal{L}\left[
  t\left\langle \overline{\delta\bm{r}^2}(\Delta; t) \right\rangle
  \right](\Delta; s)
  &
  \underset{\begin{subarray}{c}
      s\Delta \ll 1\\[.03cm] s\to 0
    \end{subarray}}{\simeq}
  \frac {e^{s \Delta} - 1}{s}
  \mathcal{L}\left[
  \left\langle \delta\bm{r}^2(t) \right\rangle
  \right](s)
  \notag\\[0.0cm]
  \label{e.etmsd(s)}
  &\underset{
    \begin{subarray}{c}
      s\Delta \ll 1\\[.03cm] s\to 0
    \end{subarray}}{\simeq}
  \Delta
  \mathcal{L}\left[
  \left\langle \delta\bm{r}^2(t) \right\rangle
  \right](s)
\end{align}
where $s \Delta \ll 1$ is resulted from $\Delta \ll t$. The Laplace inversion
gives
\begin{equation}
  \label{e.etmsd(t).final}
  \left\langle \overline{\delta\bm{r}^2}(\Delta; t) \right\rangle
  \underset{\begin{subarray}{c}
      \Delta \ll t \\[.03cm] t\to \infty
    \end{subarray}}{\simeq}
  \frac {\Delta}{t}
  \Bigl\langle \delta\bm{r}^2(t) \Bigr\rangle.
\end{equation}
Therefore, the ETMSD only shows normal diffusion; i.e., the ETMSD depends
linearly on the lag time $\Delta$. Hence, the EMSD and ETMSD coincide only if
the EMSD shows normal diffusion, otherwise they do not coincide and thus
ergodicity breaks down.

%
For the typical non-equilibrium ensembles given by Eq.~(\ref{noneq_initial}), we
simply rewrite Eq.~(\ref{e.etmsd(t).final}) as
\begin{equation}
  \label{e.etmsd(t).non-eq}
  \left\langle \overline{\delta\bm{r}^2}(\Delta; t) \right\rangle_{\mathrm{neq}}
  \underset{\begin{subarray}{c}
      \Delta \ll t \\[.03cm] t\to \infty
    \end{subarray}}{\simeq}
  \frac {\Delta}{t}
  \Bigl\langle \delta\bm{r}^2(t) \Bigr\rangle_{\mathrm{neq}}.
\end{equation}
Therefore, we can obtain the ETMSD $\langle \overline{\delta\bm{r}^2}(\Delta; t)
\rangle_{\mathrm{neq}}$ simply from the EMSD $\langle \delta\bm{r}^2(t)
\rangle_{\mathrm{neq}}$.

\subsection {Propagator}

The propagator $P_{\pm} (\bm{r}, t | \bm{w}^0) d\bm{r}$ is the probability of
finding the particle of $\pm$ state in $(\bm{r}, \bm{r}+d\bm{r})$ at time $t$,
given that the initial ensemble is $\bm{w}^0$. Integral equations for $P_{\pm}
(\bm{r}, t| \bm{w}^0)$ can be obtained in a way similar to the analysis for
CTRWs \cite{shlesinger82, bouchaud90}:
\begin{align}
  P_{\pm} (\bm{r}, t| \bm{w}^0)
  =&
  \int d\bm{r}'  G_{\pm} (\bm{r} - \bm{r}', t) (\mathcal{I}\rho_{\pm}^0)(t)
  P_{\pm} (\bm{r}',0| \bm{w}^0)
  \notag\\[.0cm]
  +\int_{0}^{t}dt' \int\!d\bm{r}' & G_{\pm} (\bm{r} - \bm{r}', t-t')
  (\mathcal{I}\rho_{\pm})(t-t') R_{\pm} (\bm{r}',t'| \bm{w}^0),
  \\[-0.0cm]
  R_{\pm} (\bm{r}, t| \bm{w}^0)
  =&
  \int d\bm{r}' G_{\mp} (\bm{r} - \bm{r}', t) \rho_{\mp}^0(t)
  P_{\mp} (\bm{r}',0 | \bm{w}^0)
  \notag\\[0.0cm]
  + \int_{0}^{t} dt' \int d\bm{r}' & G_{\mp} (\bm{r} - \bm{r}', t-t')
  \rho_{\mp}(t-t') R_{\mp} (\bm{r}',t'| \bm{w}^0),
\end{align}
where $G_{\pm} (\bm{r}, t)$ is the Green function of the diffusion process with
the diffusion constant $D_{\pm}$, and $R_{\pm} (\bm{r}, t| \bm{w}^0) d\bm{r}dt$
is the probability of which the particle reaches the domain $(\bm{r},
\bm{r}+d\bm{r})$ and becomes $\pm$ state just in the interval $(t, t+dt)$, given
that the process starts with $\bm{w}^0$.  By the Fourier and Laplace
transformations , we obtain
\begin{align}
  \label{e.P(k,s)}
  \breve{P}_{\pm} (\bm{k}, s| \bm{w}^0) =&
  \frac {1-\hat{\rho}_{\pm}^0(s_{\pm})}{s_{\pm}} \hat{P}_{\pm} (\bm{k},0| \bm{w}^0)
  \notag\\
  &+
  \frac {1-\hat{\rho}_{\pm}(s_{\pm})}{s_{\pm}} \breve{R}_{\pm} (\bm{k},s| \bm{w}^0),
  \\[0.0cm]
  \label{e.R(k,s)}
  \breve{R}_{\pm} (\bm{k}, s| \bm{w}^0) =&
  \hat{\rho}_{\mp}^0(s_{\mp})\hat{P}_{\mp} (\bm{k},0| \bm{w}^0)
  +
  \hat{\rho}_{\mp} (s_{\mp})\breve{R}_{\mp} (\bm{k},s| \bm{w}^0),
\end{align}
where $s_{\pm} := s + D_{\pm}\bm{k}^2$, and we used the fact that the Fourier
transform of $G_{\pm} (\bm{r}, t)$ is given by $\hat{G}_{\pm} (\bm{k}, t) =
e^{-D_{\pm} \bm{k}^2 t}$ \cite{hofling13}.
Here, for simplicity, we assume that the initial position is the origin
$\bm{r}(0)=0$, i.e., $P_{\pm}(\bm{r},0 | \bm{w}^0) = p^0_{\pm} \delta(\bm{r})$
and $\hat{P}_{\pm}(\bm{k},0| \bm{w}^0) = p^0_{\pm}$. Then, from
Eqs.~(\ref{e.P(k,s)}) and (\ref{e.R(k,s)}), we obtain a general expression for
the propagator:
\begin{align}
  \label{e.P(k,s).general}
  \breve{P}_{\pm}& (\bm{k}, s| \bm{w}^0) =
  \frac {1-\hat{\rho}_{\pm}^0(s_{\pm})}{s_{\pm}} p_{\pm}^0
  \notag\\[0.0cm]
  &+
  \frac {1-\hat{\rho}_{\pm}(s_{\pm})}{s_{\pm}}
  \frac
  {\hat{w}_{\pm}^0(s_{\pm}) \hat{\rho}_{\mp}(s_{\mp}) + \hat{w}_{\mp}^0(s_{\mp})}
  {1-\hat{\rho}_+(s_+) \hat{\rho}_-(s_-)}.
\end{align}


Using Eqs.~(\ref{e.eq-ensemble-1}), (\ref{e.eq-ensemble-2}) and
(\ref{e.P(k,s).general}), we have the equilibrium propagator,
$\breve{P}_{\pm}^{\mathrm{eq}} (\bm{k}, s) := \breve{P}_{\pm} (\bm{k}, s|
\bm{w}^{\mathrm{eq}})$, as
\begin{equation}
  \breve{P}_{\pm}^{\mathrm{eq}} (\bm{k}, s) =
  \frac {p_{\pm}}{s_{\pm}}
  +
  \frac
  {[1-\hat{\rho}_+(s_+)][1-\hat{\rho}_-(s_-)]}
  {\mu s_{\pm}[1-\hat{\rho}_+(s_+)\hat{\rho}_-(s_-)]}
  \left(\frac {1}{s_{\mp}} - \frac {1}{s_{\pm}}\right).
\end{equation}
Note that if $D_+=D_-$, the second term vanishes and the system shows normal
diffusion with the Gaussian propagator all the time. Therefore, the second term
is the contribution from the fluctuating diffusivity, and the propagator is
non-Gaussian in general if $D_+\neq D_-$. For the equilibrium processes, using
the asymptotic relation given in Eq.~(\ref{e.rho(s).asymptotic.alpha>1}), and
assuming that $s_{\pm}\ll 1$ along with $s \sim \bm{k}^2 \ll 1$ (i.e., a
hydrodynamic limit), we have
\begin{equation}
  \label{e.propagator.eq}
  \breve{P}_{\pm}^{\mathrm{eq}} (\bm{k}, s)
  \underset{\begin{subarray}{c}
      s, \bm{k} \to 0 \\[-.03cm]  s \sim \bm{k}^2
    \end{subarray}}{\simeq}
  \frac {p_{\pm}^{\mathrm{eq}}}{s + D_{\mathrm{eq}}\bm{k}^2}.
\end{equation}
Thus, in this hydrodynamic limit, the propagators for both $+$ and $-$ states
become Gaussian \cite{hofling13}. Note also that, the sum of the propagators,
$\breve{P}_+^{\mathrm{eq}} (\bm{k}, s) + \breve{P}_-^{\mathrm{eq}} (\bm{k}, s)$,
becomes a Gaussian distribution; this is the propagator of the diffusion process
with the diffusion constant $D_{\mathrm{eq}}$.


From Eqs. (\ref{noneq_initial}) and (\ref{e.P(k,s).general}), we have the
non-equilibrium propagator, $\breve{P}_{\pm}^{\mathrm{neq}} (\bm{k}, s) :=
\breve{P}_{\pm} (\bm{k}, s| \bm{w}^{\mathrm{neq}})$, as
\begin{equation}
  \label{e.P(k,s).general.neq}
  \breve{P}_{\pm}^{\mathrm{neq}} (\bm{k}, s)
  \underset{\begin{subarray}{c}
      s, \bm{k} \to 0 \\[-.03cm]  s \sim \bm{k}^2
    \end{subarray}}{\simeq}
  \frac 1{s_{\pm}}
  \frac
  {1-\hat{\rho}_{\pm}(s_{\pm})}
  {1-\hat{\rho}_+(s_+)\hat{\rho}_-(s_-)},
\end{equation}
where an approximation $p_{\mp}^0 \hat{\rho}_{\mp}(s_{\mp}) + p_{\pm}^0 \simeq
1$ is used. Thus, Eq.~(\ref{e.P(k,s).general.neq}) is exact only for its leading
order terms.

\subsection {Relative standard deviation of TMSD}\label{s.rsd.general}

As presented in the previous subsections, both EMSD and ETMSD exhibit normal
diffusion for equilibrium processes. Thus, with these MSDs, we cannot extract
any information on fluctuating diffusivity. To characterize such anomaly, we may
have to consider some higher order moments.  Here, we study the RSD of TMSD:
\begin{equation}
  \label{e.def.rsd}
  \Sigma(\Delta;t | \bm{w}^0) :=
  \frac{\sqrt{\bigl\langle \bigr|\overline{\delta \bm{r}^{2}}(\Delta;t)\bigr|^{2} -
      \bigl\langle \overline{\delta\bm{r}^2}(\Delta;t) \bigr\rangle^2 \bigr\rangle}}
  {\bigl\langle \overline{\delta\bm{r}^2}(\Delta;t) \bigr\rangle}.
\end{equation}
Note that $\Sigma(\Delta;t | \bm{w}^0)$ is a fourth moment of $\bm{r}(t')$. It
will be shown that, for the equilibrium process, this RSD is related to the ACF,
$\left\langle D(t)D(0)\right\rangle_{\mathrm{eq}}$ [see
Eq.~(\ref{e.RSD(t)-<D(t)D(0)>})]. The RSD analysis is useful even for the
non-equilibrium processes, because the RSD is related to non-Gaussianity of the
propagator (see Appendix \ref{s.non-gaussian}).

By assuming that $\Delta$ is much smaller than the characteristic correlation
time of $D(t)$, we can make the analysis relatively simple (This formal
approximation is represented by the symbol "$\approx$"). For example, under this
assumption, we have $\left\langle D(\tau) \right\rangle \approx \left\langle
D(t') \right\rangle$ in the integrand of Eq.~(\ref{e.etmsd(t).1}), and thus
obtain
\begin{equation}
  \label{e.etmsd.1mmt.1}
  \left\langle \overline{\delta\bm{r}^2}(\Delta; t) \right\rangle
  \approx
  \frac {2 n \Delta}{t-\Delta}\int_{0}^{t-\Delta} dt' \left\langle D(t') \right\rangle.
\end{equation}
This equation is consistent with Eqs.~(\ref{e.emsd}) and
(\ref{e.etmsd(t).final}) [though Eq.~(\ref{e.etmsd(t).final}) is valid even if
$\Delta$ is larger than the correlation time of $D(t)$]. In this subsection, we
analyze the RSD on the basis of this approximation.




With the help of Wick's theorem,
\begin{align}
  \label{e.wick}
  \sum_{i,j}\left\langle \xi_i(s)\xi_i(s')\xi_j(u)\xi_j(u')
  \right\rangle
  &= \notag\\[0.0cm]
  n^2\delta(s-s')\delta(u-u') + &2n \delta(s-u')\delta(s'-u),  
\end{align}
the second moment of the TMSD can be approximated as follows:
\begin{widetext}
  \begin{align}
    \left\langle \bigl|\overline{\delta\bm{r}^2}(\Delta; t)\bigr|^2 \right\rangle
    &=
    \frac {2}{(t-\Delta)^2} \int_{0}^{t-\Delta} dt' \int_{0}^{t'} dt''
    \left\langle
    \left[ \bm{r}(t'+\Delta) - \bm{r}(t') \right]^2
    \left[ \bm{r}(t''+\Delta) - \bm{r}(t'') \right]^2
    \right\rangle.
    \notag\\[0.0cm]
    \label{e.etmsd.2mmt.1}
    &\approx
    \frac {8n\Delta^2}{(t-\Delta)^2}
    \!\!\int_{0}^{t-\Delta} \!\!\!\!\!\!\! dt' \!\! \int_{0}^{t'} \!\!\! dt'' \!
    \left[n \left\langle D(t')D(t'') \right\rangle +
    \frac {2 (t''-t'+\Delta)^2}{\Delta^2} I(t' \leq t'' +\Delta) \left\langle D^2(t') \right\rangle\right]
    \\[0.0cm]
    \label{e.etmsd.2mmt.2}
    &=
    \frac {8n^2\Delta^2}{(t-\Delta)^2}
    \left[
    \int_{0}^{t-\Delta} dt' \int_{0}^{t'} dt''
    \left\langle D(t')D(t'') \right\rangle
    +
    \frac {2\Delta}{3n}
    \int_{0}^{t-\Delta} dt'
    \left\langle D^2(t') \right\rangle
    \right],
  \end{align}
\end{widetext}
where $I(\dots) = 1$ if the inside of the bracket is satisfied, 0 otherwise. In
deriving Eq.~(\ref{e.etmsd.2mmt.1}), we used approximations
\begin{equation}
  \int_{t'}^{t'+\Delta} ds \int_{t''}^{t''+\Delta} du
  \left\langle D(s)D(u) \right\rangle
  \approx
  \left\langle D(t')D(t'') \right\rangle \Delta^2,
\end{equation}
and $\left\langle D(s)D(u) \right\rangle \approx \left\langle D^2(t')
\right\rangle$ for $s, u \in (t'-\Delta, t'+\Delta)$. These approximations are
justified, because $\Delta$ is much smaller than the characteristic correlation
time of $D(t)$.


Inserting Eqs.~(\ref{e.etmsd.1mmt.1}) and (\ref{e.etmsd.2mmt.2}) into
Eq.~(\ref{e.def.rsd}), we obtain
\begin{equation}
  \label{e.RSD(t).general}
  \Sigma^{2} (\Delta; t | \bm{w}^0)
  \underset{\Delta \ll t}{\approx}
  \Sigma_{\mathrm{id}}^2(\Delta; t | \bm{w}^0) +
  \Sigma_{\mathrm{ex}}^2(t | \bm{w}^0),
\end{equation}
with
\begin{align}
  \label{e.RSD(t).sigma_id}
  \Sigma^{2}_{\mathrm{id}} (\Delta; t | \bm{w}^0) &:=
  \frac {4\Delta}{3n}
  \frac
  {\displaystyle\int_{0}^{t} dt' \left\langle D^2(t')  \right\rangle}
  {\displaystyle\int_{0}^{t} dt' \int_{0}^{t} dt''
    \left\langle D(t')\right\rangle\! \left\langle  D(t'') \right\rangle},
  \\[0.cm]
  \label{e.RSD(t).sigma_ex}  
  \Sigma^{2}_{\mathrm{ex}} (t | \bm{w}^0) &:=
  \frac
  {\displaystyle\int_{0}^{t} dt' \int_{0}^{t} dt''
    \left\langle \delta D(t')\delta D(t'') \right\rangle}
  {\displaystyle\int_{0}^{t} dt' \int_{0}^{t} dt''
    \left\langle D(t')\right\rangle \!\left\langle  D(t'') \right\rangle},
\end{align}
where $\delta D(t')$ is defined by $\delta D(t') := D(t') - \left\langle D(t')
\right\rangle$.
If there is no fluctuating diffusivity, then $\Sigma^{2}_{\mathrm{ex}} (t |
\bm{w}^0) = 0$ since $\delta D(t) = 0$, and thus only $\Sigma^{2}_{\mathrm{id}}
(\Delta; t | \bm{w}^0)$ contributes the RSD. Namely, $\Sigma^{2}_{\mathrm{id}}
(\Delta; t | \bm{w}^0)$ is the RSD for single-mode diffusion processes, and thus
we call this term as an ideal part. In contrast, a contribution from the
fluctuating diffusivity is represented by $\Sigma^{2}_{\mathrm{ex}} (t |
\bm{w}^0)$, an excess part of the RSD.

For the cases studied in this paper [Eqs.~(\ref{e.rho(s).asymptotic.alpha<1})
and (\ref{e.rho(s).asymptotic.alpha>1})], the leading term of the ideal part
$\Sigma_{\mathrm{id}}^2 (\Delta; t | \bm{w}^0)$ at large $t$ is estimated as
\begin{equation}
  \label{e.RSD(t).sigma_id.asympt}
  \Sigma_{\mathrm{id}}^2 (\Delta; t | \bm{w}^0)
  \underset{t \to \infty}{=}
  O\left(\frac {1}{t}\right).
\end{equation}
This can be obtained as follows. First, note that $\left\langle D^2(t)
\right\rangle = D_+^2 p_+(t|\bm{w}^0) + D_-^2 p_-(t|\bm{w}^0)$, and this is the
same as Eq.~(\ref{e.D(t).general}) except that $D_{\pm}$ is replaced by
$D_{\pm}^2$. Therefore, $2n\int_{0}^{t-\Delta} dt' \left\langle D^2(t')
\right\rangle$ has the same form as the EMSD $\left\langle \delta\bm{r}^2(t)
\right\rangle= 2n \int_{0}^{t}dt' \left\langle D(t') \right\rangle$ except that
$D_{\pm}$ is replaced by $D_{\pm}^2$. As shown in Secs.~\ref{s.emsd.general} and
\ref{s.case_study_msd}, the EMSD shows only normal diffusion $\left\langle
\delta\bm{r}^2(t) \right\rangle = O(t)$ as $t \to \infty$ [see
Eqs.~(\ref{e.emsd(t).eq}), (\ref{e.EMSD(t).case(1-1)}),
(\ref{e.EMSD(t).case(1-2)}), (\ref{e.EMSD(t).case(2-1)}),
(\ref{e.EMSD(t).case(2-2)}), (\ref{e.EMSD(t).case(3a)}), and
(\ref{e.EMSD(t).case(3b)})]. Therefore, we can conclude that
\begin{equation}
  2n\int_{0}^{t-\Delta} dt' \left\langle D^2(t') \right\rangle
  \underset{t \to \infty}{=}
  O(t)
\end{equation}
From these facts and Eq.~(\ref{e.RSD(t).sigma_id}), we obtain
(\ref{e.RSD(t).sigma_id.asympt}).

In particular, for the case (3), we shall obtain $\int_{0}^{t-\Delta} dt'
\left\langle D(t') \right\rangle = D_{\pm}t + O(t^{\alpha_{\pm}})$
[Eqs.~(\ref{e.EMSD(t).case(3a)}) and (\ref{e.EMSD(t).case(3b)})]. Thus, we have
$\int_{0}^{t-\Delta} dt' \left\langle D^2(t') \right\rangle = D_{\pm}^2t +
O(t^{\alpha_{\pm}})$, and whereby obtaining from Eq.~(\ref{e.RSD(t).sigma_id})
\begin{equation}
  \label{e.RSD(t).sigma_id.asympt.case(3)}
  \Sigma_{\mathrm{id}}^2 (\Delta; t | \bm{w}^{\mathrm{neq}})
  \underset{t \to \infty}{=}
  \frac {4\Delta}{3nt}
  +
  O \left(t^{\alpha_{\pm}-2}\right)
\end{equation}
where $\alpha_{\pm} \in (0,1)$.

Thus, the leading term of the ideal part $\Sigma^{2}_{\mathrm{id}} (\Delta; t |
\bm{w}^0)$ is of the order of $1/t$. Therefore, if the excess part
$\Sigma_{\mathrm{ex}}^2 (t | \bm{w}^0)$ shows slow relaxation $1/t^{\beta}$ with
$\beta \in (0,1)$, this part $\Sigma_{\mathrm{ex}}^2 (t | \bm{w}^0)$ is dominant
over the ideal part.  In contrast, if the excess part $\Sigma_{\mathrm{ex}}^2 (t
| \bm{w}^0)$ shows fast relaxation $1/t^{\beta}$ with $\beta > 1$, the ideal
part $\Sigma_{\mathrm{id}}^2 (\Delta; t | \bm{w}^0)$ is dominant over the excess
part. The former is actually the case for almost all the cases studied in this
article except a couple of narrow parameter regions in case (3) [see
Secs.~\ref{s.rsd.case3.1} and \ref{s.rsd.case3.2}].

\subsubsection {equilibrium processes}

For the equilibrium processes, we have from Eq.~(\ref{e.RSD(t).sigma_ex})
\begin{equation}
  \Sigma^{2}_{\mathrm{ex}} (t | \bm{w}^{\mathrm{eq}})
  \label{e.RSD(t).equilibrium}
  \simeq
  \frac {2}{t^2}
  \int_0^t dt' (t- t')
  \frac {\left\langle \delta D(t') \delta D(0)\right\rangle_{\mathrm{eq}}}{D_{\mathrm{eq}}^2},
\end{equation}
where we assumed that $\left\langle \delta D(t')\delta D(t'')
\right\rangle_{\mathrm{eq}}$ depends only on the time lag $t' - t''$ (see
Appendix \ref{s.joint.pdf.general} for a proof). Let us rewrite the above
equation as
\begin{equation}
  \label{e.RSD(t)-<D(t)D(0)>}
  \left\langle \delta D(t) \delta D(0)\right\rangle_{\mathrm{eq}}
  \underset{}{\simeq}
  \frac {D_{\mathrm{eq}}^2}{2}
  \frac {\partial^2 }{\partial t^2}
  \left[
  t^2\Sigma^{2}_{\mathrm{ex}} (t| \bm{w}^{\mathrm{eq}}) 
  \right].
\end{equation}
As shown in Sec.~\ref{s.case_study_rsd_eq}, the excess RSD,
$\Sigma^{2}_{\mathrm{ex}} (t| \bm{w}^{\mathrm{eq}})$, decays slower than $1/t$
for the case (2), and thus $\Sigma^{2} (\Delta; t| \bm{w}^{\mathrm{eq}}) \approx
\Sigma^{2}_{\mathrm{ex}} (t| \bm{w}^{\mathrm{eq}})$. Therefore,
Eq.~(\ref{e.RSD(t)-<D(t)D(0)>}) implies that we can obtain the ACF,
$\left\langle \delta D(t) \delta D(0)\right\rangle_{\mathrm{eq}}$, by measuring
the RSD, $\Sigma^{2} (\Delta; t| \bm{w}^{\mathrm{eq}})$. It is important that
the RSD and $D_{\mathrm{eq}}$ can be calculated from dozens of trajectories. By
contrast, the instantaneous DC, $D(t)$, is difficult to measure directly in
experiments.
The Laplace transform of Eq.~(\ref{e.RSD(t).equilibrium}) with respect to $t$
becomes
\begin{equation}
  \label{e.RSD(s).equilibrium}
  \mathcal{L}\left[
  t^2\Sigma^2_{\mathrm{ex}} (t| \bm{w}^{\mathrm{eq}})
  \right] (s)
  \simeq
  \frac {2}{D_{\mathrm{eq}}^2}
  \frac {\langle \delta \hat{D}(s) \delta D(0)\rangle_{\mathrm{eq}}}{s^2}
\end{equation}


Moreover, if ${\left\langle \delta D(t')\delta D(0)
  \right\rangle_{\mathrm{eq}}}$ decays faster than $O(1/t)$, we have a simple
relation from Eq.~(\ref{e.RSD(t).equilibrium}):
\begin{equation}
 \label{rsd_square_asymptotic}
 \Sigma^{2}_{\mathrm{ex}}(t| \bm{w}^{\mathrm{eq}})
 \underset{
   \begin{subarray}{c}
     t \to \infty
   \end{subarray}
   }{\simeq}
 \frac{2}{t} \int_{0}^{\infty} dt'
 \frac
 {\left\langle \delta D(t')\delta D(0) \right\rangle_{\mathrm{eq}}}
 {D_{\mathrm{eq}}^2}.
\end{equation}
This type of decay is observed in the center of mass motion of the reptation
model \cite{uneyama12, uneyama15}. But, if the ACF, ${\left\langle \delta D(t')\delta D(0)
  \right\rangle_{\mathrm{eq}}}$, decays slower than $O(1/t)$, the
Eq.~(\ref{rsd_square_asymptotic}) is no more valid, since the integral
diverges. In such cases, the RSD exhibits slow relaxation as shown in
Sec.~\ref{s.case_study_rsd_eq}.


\subsubsection {Non-equilibrium processes}
For non-equilibrium processes, a simple relation between the RSD and the ACF
$\left\langle \delta D(t_1) \delta D(t_2)\right\rangle$ [such as the
Eq.~(\ref{e.RSD(t)-<D(t)D(0)>})] does not hold.
Instead of the RSD, we define
\begin{equation}
  \label{e.RSD1(t1,t2).non-equilibrium}
  \Sigma^2_{\mathrm{ex}} (t_1, t_2| \bm{w}^{\mathrm{neq}}):=
  \frac
  {\displaystyle\int_{0}^{t_1} dt' \int_{0}^{t_2} dt''
    \left\langle \delta D(t')\delta D(t'')
    \right\rangle_{\mathrm{neq}}}
  {\displaystyle\int_{0}^{t_1} dt' \int_{0}^{t_2} dt''
    \left\langle D(t')  \right\rangle_{\mathrm{neq}}
    \left\langle D(t'') \right\rangle_{\mathrm{neq}}},
\end{equation}
then $\Sigma^2_{\mathrm{ex}} (t| \bm{w}^{\mathrm{neq}}) =
\Sigma^2_{\mathrm{ex}} (t, t| \bm{w}^{\mathrm{neq}})$. From
Eqs.~(\ref{e.emsd}) and (\ref{e.RSD1(t1,t2).non-equilibrium}), we have
\begin{align}
  \label{e.RSD(s).non-equilibrium}
  \mathcal{L}^2\Bigl[
  \left\langle \delta\bm{r}^2(t_1) \right\rangle_{\mathrm{neq}}&
  \left\langle \delta\bm{r}^2(t_2) \right\rangle_{\mathrm{neq}}
  \Sigma^2_{\mathrm{ex}} (t_1, t_2| \bm{w}^{\mathrm{neq}})
  \Bigr] (s_1, s_2)\notag\\[0.0cm]
  &=
  4n^2
  \frac {\langle \delta \hat{D}(s_1) \delta \hat{D}(s_2)\rangle_{\mathrm{neq}}}{s_1s_2},
\end{align}
where $\mathcal{L}^2$ stands for double Laplace transforms with respect to
$t_1$ and $t_2$.

\subsection {Auto-correlation function of diffusion coefficient}

Here, we study general expressions for the ACF $\left\langle D(t_1)D(t_2)
\right\rangle$, from which the excess RSD can be analytically obtained
[Eqs.~(\ref{e.RSD(t).equilibrium}) and (\ref{e.RSD1(t1,t2).non-equilibrium})].
For the equilibrium processes, we have $\left\langle \delta D(t) \delta
D(0)\right\rangle_{\mathrm{eq}} = \left\langle D(t)
D(0)\right\rangle_{\mathrm{eq}} - D_{\mathrm{eq}}^2$, where the first term is
expressed with the transition probabilities in Eq.~(\ref{e.W(s).eq}) as
\begin{equation}
  \left\langle D(t) D(0)\right\rangle_{\mathrm{eq}}
  =
  \sum_{h,h' = \pm}^{} D_hD_{h'} W_{hh'}^{\mathrm{eq}} (t),
\end{equation}
and the Laplace transform is given by
\begin{equation}
  \label{e.<D(s)D(0)>}
  \langle \hat{D}(s) D(0) \rangle_{\mathrm{eq}}
=
\sum_{h,h' = \pm}^{} D_hD_{h'} \hat{W}_{hh'}^{\mathrm{eq}} (s).
\end{equation}
From Eqs.~(\ref{e.RSD(s).equilibrium}) and (\ref{e.<D(s)D(0)>}), we can obtain asymptotic
behavior of the RSD.


For the non-equilibrium ensembles, we have to consider the ACF $\left\langle
D(t_1)D(t_2) \right\rangle$.  Using a general transition probability,
$W_{hh'}^{\mathrm{neq}} (t_1, t_2)$, which is a joint probability that the state
is $h$ at time $t_1$ and $h'$ at time $t_2$, given that the initial ensemble is
$\bm{w}^{\mathrm{neq}}$ (see Appendix \ref{s.joint.pdf.general}), we have
\begin{equation}
  \label{e.<D(t')D(t'+t)>}
  \!\left\langle D(t_1)D(t_2) \right\rangle_{\mathrm{neq}}
  \!=\!\!\!
  \sum_{h,h' = \pm}^{} \!\!\! D_hD_{h'}
  \left[
  W_{hh'}^{\mathrm{neq}} (t_1, t_2) + W_{h'h}^{\mathrm{neq}} (t_2, t_1)  
  \right],
\end{equation}
where the second term $W_{h'h}^{\mathrm{neq}} (t_2, t_1)$ is necessary because
$W_{hh'} (t_1,t_2) = 0$ if $t_2<t_1$ from its definition (see Appendix
\ref{s.joint.pdf.general}), whereas the ACF, $\left\langle D(t_1)D(t_2)
\right\rangle$, is defined for both $t_1 < t_2$ and $t_2 < t_1$.  Then, the
Laplace transforms  with respect to $t_1$ and
$t_2$ result in
\begin{equation}
  \!\langle \hat{D}(s_1) \hat{D}(s_2) \rangle_{\mathrm{neq}}
  \!=\!\!\!
  \sum_{h,h' = \pm}^{}\!\!\!D_hD_{h'}\!
  \left[
  \breve{W}_{hh'}^{\mathrm{neq}} (s_1, s_2) +
  \breve{W}_{hh'}^{\mathrm{neq}} (s_2, s_1)
  \right],
\end{equation}
where $\breve{W}_{hh'}^{\mathrm{neq}} (s_1, s_2)$ is given by
Eq.~(\ref{e.W(s1,s2).neq}).

Moreover, let us define $\delta W_{hh'}^{\mathrm{neq}}(t_1, t_2)$ as
\begin{equation}
  \label{e.def.delta_W(t1,t2)}
  \!\delta W_{hh'}^{\mathrm{neq}}(t_1, t_2) :=
  W_{hh'}^{\mathrm{neq}}(t_1, t_2) + W_{h'h}^{\mathrm{neq}}(t_2, t_1)
  -
  p_{h}^{\mathrm{neq}}(t_1)p_{h'}^{\mathrm{neq}}(t_2),
\end{equation}
where $p_{h}^{\mathrm{neq}}(t)$ is the fraction of the state $h$ at time $t$.
Then, the ACF of deviations from the mean values, $\langle \delta\hat{D}(s_1)
\delta\hat{D}(s_2) \rangle_{\mathrm{neq}}$, can be expressed as
\begin{equation}
  \label{e.<dD(s1)dD(s2)>}
  \langle \delta\hat{D}(s_1) \delta\hat{D}(s_2) \rangle_{\mathrm{neq}}
  =\!\!
  \sum_{h,h' = \pm}^{}\!\!D_hD_{h'}
  \delta\breve{W}_{hh'}^{\mathrm{neq}} (s_1, s_2).
\end{equation}
From Eqs.~(\ref{e.RSD(s).non-equilibrium}) and (\ref{e.<dD(s1)dD(s2)>}), we can
obtain asymptotic behavior of the non-equilibrium RSD.

\section {Case study: fractions of states}\label{s.case_study_fraction}

In this and subsequent sections, we study special cases in which the sojourn
time PDFs, $\rho_{\pm}(\tau)$, are given by the power laws defined in
Sec.~\ref{s.two-state-system}.  In this section, we investigate only the
non-equilibrium fraction of states, $p_{\pm}^{\mathrm{neq}}(t)$, since the
equilibrium fractions are constant, $p_{\pm}^{\mathrm{eq}}(t) \equiv
p_{\pm}^{\mathrm{eq}}$ [Sec.~\ref{s.general.fraction}].  It is shown that the
asymptotic behaviors of $p_{\pm}^{\mathrm{neq}}(t)$ are independent of the
initial fractions $p_{\pm}^0$.


\subsection {Fractions for case (1)}

Let us start with the case (1).  For $\alpha_+ < \alpha_-$ [case (1-1)], we have
from Eqs.~(\ref{e.rho(s).asymptotic.alpha<1}) and (\ref{e.fraction.p(s).noneq})
\begin{equation}
  \label{e.p(s).case(1-1)}
  \hat{p}_+^{\mathrm{neq}} (s)
  \underset{s \to 0}{\simeq}
  \frac {1}{s} -
  \frac {a_-}{a_+} s^{\alpha_- - \alpha_+ - 1},
\end{equation}
and the Laplace inversion is given by
\begin{equation}
  \label{e.p(t).case(1-1)}
  p_+^{\mathrm{neq}} (t)
  \underset{t \to \infty}{\simeq}
  1 - \frac {a_-}{a_+}
  \frac {t^{\alpha_+ - \alpha_-}}{\Gamma(\alpha_+ - \alpha_- + 1)}.
\end{equation}
Thus, the ensemble accumulates to the $+$ state for large $t$. On the other
hand, $p_-^{\mathrm{neq}}(t)$ is given simply by $p_-^{\mathrm{neq}}(t) = 1 -
p_+^{\mathrm{neq}}(t)$. Note that the dependence on the initial fractions
$p_{\pm}^0$ vanishes in Eq.~(\ref{e.p(t).case(1-1)}), and thus the asymptotic
behavior is independent of the initial fractions, $p^0_{\pm}$.  From
Eq.~(\ref{e.D(t).general.neq}), the mean DC is given by
\begin{equation}
  \label{e.D(t).case(1-1)}
  \left\langle D(t) \right\rangle_{\mathrm{neq}}
  \underset{t \to \infty}{\simeq}
  D_+
  -
  \frac {a_-}{a_+}
  \frac {(D_+-D_-)\,t^{\alpha_+ - \alpha_-}}{\Gamma(\alpha_+ - \alpha_- + 1)}
\end{equation}
and therefore $\left\langle D(t) \right\rangle_{\mathrm{neq}}$ coverages to
$D_+$ as $t \to \infty$.

Similarly, for $\alpha_- < \alpha_+$ [case (1-2)], we have
\begin{align}
  \label{e.p(s).case(1-2)}
  \hat{p}_+^{\mathrm{neq}} (s)
  &\underset{s \to 0}{\simeq}
  \frac {a_+}{a_-} s^{\alpha_+ - \alpha_- - 1},
  \\[0.0cm]
  \label{e.p(t).case(1-2)}
  p_+^{\mathrm{neq}} (t)
  &\underset{t \to \infty}{\simeq}
  \frac {a_+}{a_-} \frac {t^{\alpha_- - \alpha_+}}{\Gamma(\alpha_- - \alpha_+ + 1)}.
\end{align}
Consequently,
the mean DC is given by
\begin{equation}
  \label{e.D(t).case(1-2)}
  \left\langle D(t) \right\rangle_{\mathrm{neq}}
  \underset{t \to \infty}{\simeq}
  D_- + 
  \frac {a_+}{a_-}
  \frac {(D_+-D_-) t^{\alpha_- - \alpha_+}}
  {\Gamma(\alpha_- - \alpha_+ + 1)},
\end{equation}
and therefore $\left\langle D(t) \right\rangle_{\mathrm{neq}}$ converges to
$D_-$ as $t \to \infty$.

』

The asymptotic equations (\ref{e.p(t).case(1-1)}) and (\ref{e.D(t).case(1-1)})
[case (1-1): $\alpha_+< \alpha_-$] are valid for
\begin{equation}
  \label{e.p(t).case(1-1).range}
  \frac {a_-}{a_+} t^{\alpha_+ - \alpha_-} \ll 1,
\end{equation}
and, similarly, Eqs.~(\ref{e.p(t).case(1-2)}) and (\ref{e.D(t).case(1-2)}) [case
(1-2): $\alpha_-< \alpha_+$] are valid for
\begin{equation}
  \label{e.p(t).case(1-2).range}
  \frac {a_+}{a_-} t^{\alpha_- - \alpha_+} \ll 1.
\end{equation}
Note that these inequalities are only necessary conditions. To obtain a
sufficient condition, we need to specify higher order terms, $O(s)$, of
$\hat{\rho}_{\pm}(s)$ in Eq.~(\ref{e.rho(s).asymptotic.alpha<1}) more
concretely. As shown below, however, the conditions,
Eqs.~(\ref{e.p(t).case(1-1).range}) and (\ref{e.p(t).case(1-2).range}), are
useful to analyze crossover phenomena appearing in the EMSD.

\subsection {Fractions for case (2)}

For the case (2), the equilibrium state exists, and any non-equilibrium state
relaxes to the equilibrium as shown in Eq.~(\ref{e.p(t)->p^eq(t)}).
For the non-equilibrium ensemble [Eq.~(\ref{noneq_initial})], we have from
Eqs.~(\ref{e.rho(s).asymptotic.alpha>1}) and (\ref{e.fraction.p(s).noneq})
\begin{align}
  \label{e.p(s).case(2-1)}
  \hat{p}_+^{\mathrm{neq}} (s)
  &\underset{s \to 0}{\simeq}
  \frac {p_+^{\mathrm{eq}}}{s}
  -
  \frac {
    p_-^{\mathrm{eq}} a_+ 
  }{\mu s^{2-\alpha_+}}
  \\[0.0cm]
  \label{e.p(t).case(2-1)}
  p_+^{\mathrm{neq}} (t)
  &\underset{t \to \infty}{\simeq}
  p_+^{\mathrm{eq}}
  -
  \frac {p_-^{\mathrm{eq}} a_+ t^{1-\alpha_+}}{\mu\Gamma(2-\alpha_+)}. 
\end{align}
for $\alpha_+ < \alpha_-$ [case (2-1)].
Thus, the non-equilibrium fractions $p_{\pm}^{\mathrm{neq}} (t)$ converge to the
equilibrium ones $p_{\pm}^{\mathrm{eq}}$.
Similarly, for $\alpha_- < \alpha_+$ [case (2-2)], we have
\begin{align}
  \label{e.p(s).case(2-2)}
  \hat{p}_+^{\mathrm{neq}} (s)
  &\underset{s \to 0}{\simeq}
  \frac {p_+^{\mathrm{eq}}}{s}
  +
  \frac {
  p_+^{\mathrm{eq}} a_- 
  }{\mu s^{2- \alpha_-}},
  \\[0.cm]
  \label{e.p(t).case(2-2)}
  p_+^{\mathrm{neq}} (t)
  &\underset{t \to \infty}{\simeq}
  p_{+}^{\mathrm{eq}}
  +
  \frac {p_+^{\mathrm{eq}} a_- t^{1-\alpha_-}}{\mu\Gamma(2-\alpha_-)}. 
\end{align}
%
%
Thus, in both cases, the fractions $p_{\pm}^{\mathrm{neq}}(t)$ converge to the
equilibrium ones $p_{\pm}^{\mathrm{eq}}$.
A necessary condition for the formulas given by Eqs.~(\ref{e.p(t).case(2-1)})
and (\ref{e.p(t).case(2-2)})
to be valid is
\begin{equation}
  \label{e.p(t).case(2).range}
  a_{\pm} t^{1-\alpha_{\pm}} \ll \mu_{\pm}.
\end{equation}

\subsection {Fractions for case (3)}

Finally, we derive the non-equilibrium fractions $p_{\pm}^{\mathrm{neq}} (t)$
for the case (3). For the case (3-1), we have from
Eqs.~(\ref{e.rho(s).asymptotic.alpha<1}), (\ref{e.rho(s).asymptotic.alpha>1})
and (\ref{e.fraction.p(s).noneq}) \footnote{In these cases, the $O(s)$ term in
  Eq.~(\ref{e.rho(s).asymptotic.alpha<1}) should be treated precisely by
  assuming $\hat{\rho}_{\pm}(s) \simeq 1 - a_{\pm} s^{\alpha_{\pm}} +
  b_{\pm}s$. But, $b_{\pm}$ does not appear in the final results
  [Eqs.~(\ref{e.p(s).case(3a)})--(\ref{e.p(t).case(3b)}) or
  Eqs.~(\ref{e.delta.Whh(s1,s2).case(3a-1)}) and
  (\ref{e.delta.Whh(s1,s2).case(3a-2)})].}
\begin{equation}
  \label{e.p(s).case(3a)}
  \hat{p}_+^{\mathrm{neq}} (s)
  \underset{s \to 0}{=}
  \frac {1}{s} - \frac {\mu_-}{a_+} s^{-\alpha_+} + o(s^{-\alpha_+}),
\end{equation}
where the leading contribution from the higher order term $o(s^{-\alpha_+})$ is
given by \footnote{This higher order term can be easily derived from the
  asymptotic expansion of $\hat{p}_-^{\mathrm{neq}} (s)$ instead of
  $\hat{p}_+^{\mathrm{neq}} (s)$, for which a lengthier calculation is
  needed. This term is used only in the derivation of
  Eqs.~(\ref{e.delta.Whh(s1,s2).case(3a-1)}) and
  (\ref{e.delta.Whh(s1,s2).case(3a-2)}).}
\begin{equation}
  \label{e.p(s).case(3a)-higher-order-terms}
  \frac {\mu_-}{a_+}s^{-\alpha_+}
  \max\left(
  \frac {a_-}{\mu_-}s^{\alpha_--1},
  p_+^0 a_+ s^{\alpha_+},
  \frac {\mu_- - b_+}{a_+} s^{1-\alpha_+}
  \right).
\end{equation}
The Laplace inversion of Eq.~(\ref{e.p(s).case(3a)}) gives
\begin{equation}
  \label{e.p(t).case(3a)}
  p_+^{\mathrm{neq}} (t)
  \underset{t \to \infty}{\simeq}
  1 -  \frac {\mu_-}{a_+} \frac {t^{\alpha_+ - 1}}{\Gamma(\alpha_+)}.
\end{equation}
Similarly, for the case (3-2), we obtain
\begin{equation}
  \hat{p}_+^{\mathrm{neq}} (s)
  \underset{s \to 0}{\simeq}
  \frac {\mu_+}{a_-} s^{-\alpha_-}, \quad
  \label{e.p(t).case(3b)}
  p_+^{\mathrm{neq}} (t)
  \underset{t \to \infty}{\simeq}
  \frac {\mu_+}{a_-} \frac {t^{\alpha_- - 1}}{\Gamma(\alpha_-)}.
\end{equation}
Necessary conditions that Eqs.~(\ref{e.p(t).case(3a)}) and
(\ref{e.p(t).case(3b)}) are valid are given by
\begin{equation}
  \label{e.p(t).case(3).range}
  \frac {\mu_-}{a_+} t^{\alpha_+ - 1} \ll 1
  \quad \text{and} \quad
  \frac {\mu_+}{a_-} t^{\alpha_- - 1} \ll 1,
\end{equation}
respectively.

\section {Case study: EMSD and ETMSD}\label{s.case_study_msd}

In this section, we study the EMSD and ETMSD for the power-law sojourn time PDFs.
In contrast to the equilibrium MSDs, which behave normally
[Eqs.~(\ref{e.emsd(t).eq}) and (\ref{e.etmsd(t).eq})], the non-equilibrium EMSD
in the cases (1-2) and (3-2) [the slow-mode-dominated cases, in which $\alpha_-
< \alpha_+$] exhibits transient subdiffusion. In addition, though the
non-equilibrium ETMSD shows normal diffusion [Eq.~(\ref{e.etmsd(t).non-eq})],
transient aging is observed in an effective DC for the slow-mode-dominated
cases. A crossover time $t_c$ between these transient and asymptotic behaviors
is also derived. As with the fractions $p_{\pm}^{\mathrm{neq}}(t)$, the
asymptotic properties are independent of the initial fractions $p_{\pm}^0$.

\subsection {MSDs for case (1)}

\begin{figure}[]
  \centerline{\includegraphics[width=5.6cm]{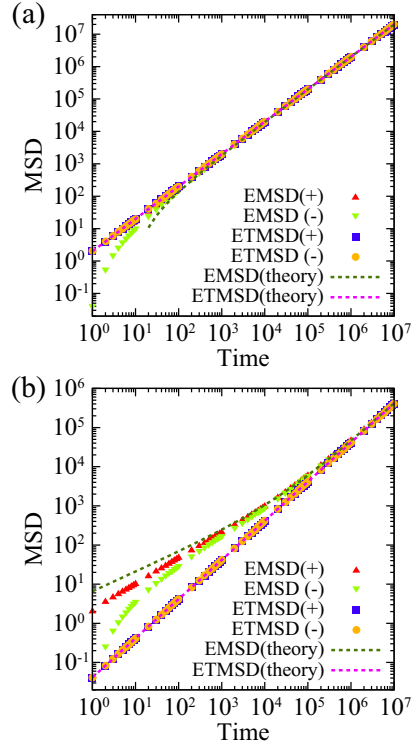}}
  \caption{\label{f.MSD_case1}Non-equilibrium EMSD $\left\langle \delta \bm{r}^2
    (\Delta) \right\rangle_{\mathrm{neq}}$ and ETMSD $\langle
    \overline{\delta\bm{r}^2}(\Delta; t) \rangle_{\mathrm{neq}}$ vs lag time
    $\Delta$ for the case (1). The total measurement time $t$ for the ETMSD
    $\langle \overline{\delta\bm{r}^2}(\Delta; t)\rangle_{\mathrm{neq}}$ is
    fixed at $t=10^7$. Two non-equilibrium ensembles are used; one is started
    from $+$ state ($p_+^0=1$: triangle up and square), while the other from $-$
    state ($p_+^0=0$: triangle down and circle). Distance is measured in units
    of $\sqrt{\mathrm{D_+} \tau_0}$ and time in units of $\tau_0$ [$\tau_0 =
    (c_{\pm}/\alpha_{\pm})^{1/\alpha_{\pm}}$ is a short-time cutoff of the
    sojourn time PDFs $\rho_{\pm}(\tau)$. See Appendix
    \ref{sec:app.simulation}]. The ratio $D_+/D_-$ is set as $D_+/D_- = 50$. (a)
    [Case (1-1)] The exponents $\alpha_{\pm}$ are set as $(\alpha_+, \alpha_-) =
    (0.25, 0.75)$. The lines are theoretical predictions given by
    Eqs.~(\ref{e.EMSD(t).case(1-1)}) and (\ref{e.ETMSD(t).case(1-1)}). (b) [Case
    (1-2)] $\alpha_{\pm}$ are set as $(\alpha_+, \alpha_-) = (0.75, 0.25)$. The
    lines are theoretical predictions given by Eqs.~(\ref{e.EMSD(t).case(1-2)})
    and (\ref{e.ETMSD(t).case(1-2)}). The crossover time $t_c$ estimated with
    Eq.~(\ref{e.EMSD.case(1-2).t_c}) is about $2\times 10^4$. There are no
    fitting parameters for both figures (The same is true of
    Figs.~\ref{f.MSD_case2} and \ref{f.MSD_case3}).}
\end{figure}


For $\alpha_+ < \alpha_-$ [case (1-1)], inserting Eq.~(\ref{e.p(s).case(1-1)})
into Eq.~(\ref{e.EMSD.general.noneq}) and performing the Laplace inversion, we
have
\begin{equation}
  \label{e.EMSD(t).case(1-1)}
  \left\langle  \delta\bm{r}^2(t) \right\rangle_{\mathrm{neq}}
  \underset{t \to \infty}{\simeq}
  2n
  \left[
  D_+ t - \frac {a_-}{a_+}
  \frac
  {(D_+ - D_-) t^{\alpha_+ - \alpha_- + 1}}
  {\Gamma (\alpha_+ - \alpha_- + 2)}
  \right].
\end{equation}
In spite of the appearance of the term $t^{\alpha_+ - \alpha_- + 1}$, transient
subdiffusion cannot be observed in this case, and thus the EMSD behaves
normally. This is because the first term is bigger than the second due to the
inequality in Eq.~(\ref{e.p(t).case(1-1).range}) and $D_+ > D_+ - D_-$ [note
also that $\Gamma (\alpha_+ - \alpha_- + 2)$ is of the order of $1$].

Similarly, for $\alpha_- < \alpha_+$ [case (1-2)], inserting
Eq.~(\ref{e.p(s).case(1-2)}) into Eq.~(\ref{e.EMSD.general.noneq}), we obtain
\begin{equation}
  \label{e.EMSD(t).case(1-2)}
  \left\langle  \delta\bm{r}^2(t) \right\rangle_{\mathrm{neq}}
  \underset{t \to \infty}{\simeq}
  2n
  \left[
  D_- t + \frac {a_+}{a_-}
  \frac
  {(D_+ - D_-) t^{\alpha_- - \alpha_+ + 1}}
  {\Gamma (\alpha_- - \alpha_+ + 2)}
  \right].
\end{equation}
In contrast to the case (1-1), transient subdiffusion can be observed, if $D_-
\ll D_+$. In fact, the crossover time $t_c$ from subdiffusion to normal
diffusion is given by
\begin{equation}
  \label{e.EMSD.case(1-2).t_c}
  \frac {a_+}{a_-} t_c^{\alpha_- - \alpha_+} \sim \frac {D_-}{D_+ - D_-},
\end{equation}
where we omitted the gamma function for simplicity, since $\Gamma(\alpha_- -
\alpha_+ +2) \sim 1$.  If $t_c$ satisfies the inequality in
Eq.~(\ref{e.p(t).case(1-2).range}), i.e., if $D_-/(D_+ - D_-) \ll 1$, we can
observe the transient subdiffusion; note that this condition is equivalent to
$D_- \ll D_+$. In particular, if $D_-=0$, the crossover time $t_c$ becomes
infinite; that is, the regime of normal diffusion vanishes and the EMSD shows
permanent subdiffusion. Therefore, if $D_-=0$, the LEFD behaves similarly to the
CTRW, which also shows permanent subdiffusion. Nevertheless, the EMSD exponents
are different between the two models; for the CTRW, the EMSD behaves as
$t^{-\alpha_-}$, whereas the present system behaves as $t^{\alpha_- - \alpha_+
  +1}$. As stated below, however, the EMSD exponent for the case (3-2) is
equivalent to that of CTRW.



Using Eqs.~(\ref{e.etmsd(t).non-eq}) and (\ref{e.EMSD(t).case(1-1)}), we can
obtain an effective DC of the ETMSD:
\begin{equation}
  \label{e.ETMSD(t).case(1-1)}
  \frac
  {\bigl\langle  \overline{\delta\bm{r}^2}(\Delta; t) \bigr\rangle_{\mathrm{neq}}}
  {2n \Delta} 
  \underset{\begin{subarray}{c}
      \Delta \ll t \\[.03cm] t\to \infty
      \end{subarray}}{\simeq} D_+ - \frac {a_-}{a_+} \frac {(D_+ -
      D_-)t^{\alpha_+ - \alpha_- }}{\Gamma (\alpha_+ - \alpha_- + 2)},
\end{equation}
for $\alpha_+ < \alpha_-$ [case (1-1)]. Thus, the effective DC increases with
the measurement time $t$; namely, the system shows relaxation to the fast state
$D_+$. However, it is hard to observe this relaxational behavior (i.e., there is
no crossover), because the first term is dominant over the second all the time
due to the inequality in Eq.~(\ref{e.p(t).case(1-1).range}) and $D_+ > D_+ -
D_-$.

Similarly, we have the ETMSD for $\alpha_- < \alpha_+$ [case (1-2)] by using
Eqs.~(\ref{e.etmsd(t).non-eq}) and (\ref{e.EMSD(t).case(1-2)}) as
\begin{equation}
  \label{e.ETMSD(t).case(1-2)}
  \frac
  {\bigl\langle  \overline{\delta\bm{r}^2}(\Delta; t) \bigr\rangle_{\mathrm{neq}}}
  {2n\Delta} 
  \underset{\begin{subarray}{c}
      \Delta \ll t \\[.03cm] t\to \infty
    \end{subarray}}{\simeq}
  D_- +  \frac {a_+}{a_-}
  \frac {(D_+ - D_-)t^{\alpha_- - \alpha_+}}{\Gamma (\alpha_- - \alpha_+ + 2)}.
\end{equation}
Again, the effective DC shows relaxation, but in this case it
decreases with time. More precisely, the effective DC shows crossover between
relaxation ($t^{\alpha_- - \alpha_+}$) at short time to a plateau ($D_-$) at
longer time; the crossover time is again given by
Eq.~(\ref{e.EMSD.case(1-2).t_c}). Thus, in contrast to the case (1-1), we can
observe relaxation of the effective DC in this case.  Moreover, if $D_- = 0$,
the effective DC tends to $0$ as $t\to \infty$. This is also a famous property
of the CTRW referred to as aging \cite{he08, barkai03a, schulz14}, though the
exponent of the present system ($\alpha_- - \alpha_+$) is different from that of
CTRW ($\alpha_- - 1$). Therefore, we call the relaxational behavior observed for $D_-
> 0$ as transient aging.

The theoretical predictions for the EMSD [Eqs.~(\ref{e.EMSD(t).case(1-1)}) and
(\ref{e.EMSD(t).case(1-2)})] and the ETMSD [Eqs.~(\ref{e.ETMSD(t).case(1-1)})
and (\ref{e.ETMSD(t).case(1-2)})] are confirmed by numerical simulations in
Fig.~\ref{f.MSD_case1}. As predicted, transient subdiffusion in the EMSD is
observed only for $\alpha_- < \alpha_+$ [Fig.~\ref{f.MSD_case1}(b)]. In
Fig.~\ref{f.MSD_case1}(a) and (b), the deflections of upside-down triangles from
the normal diffusion and transient subdiffusion, respectively, are due to the
effect of the initial ensemble; this is a higher order contribution and
neglected in the theory [similar deflections appear in cases (2) and (3) below].
Moreover, in both cases (1-1) and (1-2), the ETMSD shows only normal diffusion.

\subsection {MSDs for case (2)}

\begin{figure}[t!]
  \includegraphics[width=56 mm]{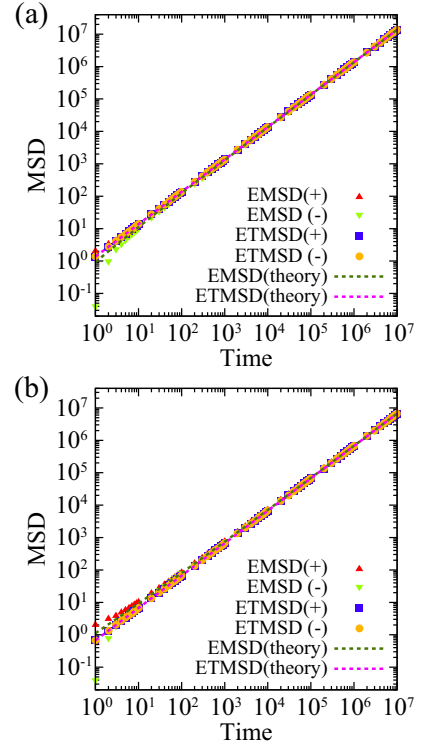}
  \caption{\label{f.MSD_case2} Non-equilibrium EMSD $\left\langle \delta
    \bm{r}^2 (\Delta) \right\rangle_{\mathrm{neq}}$ and ETMSD $\langle
    \overline{\delta\bm{r}^2}(\Delta; t) \rangle_{\mathrm{neq}}$ 
    vs lag time $\Delta$ for the case (2).  The total measurement time $t$,
    the initial fraction $p^0_+$, the distance and time units, and the ratio 
    $D_+/D_-$ are the same as in
    Fig.~\ref{f.MSD_case1}.
    (a) [Case (2-1)] $\alpha_{\pm}$ are set as $(\alpha_+, \alpha_-) = (1.25,
    1.75)$. The lines are theoretical predictions given by
    Eqs.~(\ref{e.EMSD(t).case(2-1)}) and (\ref{e.ETMSD(t).case(2-1)}). (b) [Case
    (2-2)] $\alpha_{\pm}$ are set as $(\alpha_+, \alpha_-) = (1.75, 1.25)$. The
    lines are theoretical predictions given by Eqs.~(\ref{e.EMSD(t).case(2-2)})
    and (\ref{e.ETMSD(t).case(2-2)}). }
\end{figure}

For $\alpha_+ < \alpha_-$ [case (2-1)], putting Eq.~(\ref{e.p(s).case(2-1)})
into Eq.~(\ref{e.EMSD.general.noneq}) and performing the Laplace inversion, we
have
\begin{equation}
  \label{e.EMSD(t).case(2-1)}
  \left\langle  \delta\bm{r}^2(t) \right\rangle_{\mathrm{neq}}
  \underset{t \to \infty}{\simeq}
  2n
  \left[
  D_{\rm eq} t 
  -
  (D_+ - D_-)
  \frac{a_+ p_-^{\rm eq} t^{2-\alpha_+}}{\mu\Gamma(3-\alpha_+)}
  \right].
\end{equation}
Similarly, for $\alpha_- < \alpha_+$ [case (2-2)], inserting
Eq.~(\ref{e.p(s).case(2-2)}) into Eq.~(\ref{e.EMSD.general.noneq}), we obtain
\begin{equation}
  \label{e.EMSD(t).case(2-2)}
  \left\langle  \delta\bm{r}^2(t) \right\rangle_{\mathrm{neq}}
  \underset{t \to \infty}{\simeq}
  2n \left[
  D_{\rm eq} t 
  +
  (D_+ - D_-)
  \frac{a_- p_+^{\rm eq} t^{2-\alpha_-}}{\mu\Gamma(3-\alpha_-)}
  \right].
\end{equation}
Here, the second terms in
Eqs.~(\ref{e.EMSD(t).case(2-1)})--(\ref{e.EMSD(t).case(2-2)}) are never dominant
over the first terms, and thus the only normal diffusion is observed in the
above three cases. We prove this for Eq.~(\ref{e.EMSD(t).case(2-2)}); it is also
possible to prove it for Eq.~(\ref{e.EMSD(t).case(2-1)}) 
in a similar way.

First, if $D_+ \sim D_-$, then $D_{\mathrm{eq}} \gtrsim D_+ - D_-$; from this
relation and the inequalities given Eq.~(\ref{e.p(t).case(2).range}), we can
conclude that the first term in Eq.~(\ref{e.EMSD(t).case(2-2)}) is dominant over
the second.
If $D_+ \gg D_-$, the same conclusion can be obtained as follows.  Let $t_c$ be
the crossover time between the first and second terms in
Eq.~(\ref{e.EMSD(t).case(2-2)}). If $t_c$ satisfies the condition in
Eq.~(\ref{e.p(t).case(2).range}), we can observe the crossover. This condition
becomes
\begin{equation}
  a_-t_c^{1-\alpha_-} \sim
  \frac {D_{\mathrm{eq}}}{D_+} \frac {\mu}{p_+^{\mathrm{eq}}}
  \ll
  \mu_-,
\end{equation}
where we used $D_{+} - D_{-} \sim D_+$ since $D_+ \gg D_-$ and omitted the gamma
function for simplicity, since $\Gamma(3 - \alpha_-) \sim 1$. The above
condition is rewritten as $D_-p_-^{\mathrm{eq}} \ll - D_+(p_+^{\mathrm{eq}})^2$,
which is impossible to be satisfied. Therefore, the crossover cannot be
observed, and hence the EMSD does not show the transient subdiffusion.


By using Eqs.~(\ref{e.etmsd(t).non-eq}) and (\ref{e.EMSD(t).case(2-1)}), we can
obtain the ETMSD for $\alpha_+ < \alpha_-$ as follows:
\begin{equation}
  \label{e.ETMSD(t).case(2-1)}
  \!
  \frac
  {\bigl\langle \overline{\delta\bm{r}^2}(\Delta; t) \bigr\rangle_{\mathrm{neq}}}
  {2n \Delta}
  \underset{\begin{subarray}{c}
      \Delta \ll t \\[.03cm] t\to \infty
    \end{subarray}}{\simeq}
  D_{\rm eq} 
  -
  (D_+ - D_-)
  \frac{a_+ p_-^{\rm eq} t^{1-\alpha_+}}{\mu\Gamma(3-\alpha_+)}.
\end{equation}
Thus, the effective DC becomes faster as the measurement time $t$ increases. The
opposite is the case for $\alpha_- < \alpha_+$; namely, the effective DC becomes
slower. In fact, by using Eqs.~(\ref{e.etmsd(t).non-eq}) and
(\ref{e.EMSD(t).case(2-2)}) we have
\begin{equation}
  \label{e.ETMSD(t).case(2-2)}
  \!
  \frac
  {\bigl\langle \overline{\delta\bm{r}^2}(\Delta; t) \bigr\rangle_{\mathrm{neq}}}
  {2n\Delta}
  \underset{\begin{subarray}{c}
      \Delta \ll t \\[.03cm] t\to \infty
    \end{subarray}}{\simeq}
  D_{\rm eq} 
  +
  (D_+ - D_-)
  \frac{a_- p_+^{\rm eq} t^{1-\alpha_-}}{\mu\Gamma(3-\alpha_-)}.
\end{equation}
These relaxational behaviors of the effective DC in
Eqs.~(\ref{e.ETMSD(t).case(2-1)}) and (\ref{e.ETMSD(t).case(2-2)}),
should be difficult to observe, since
$D_{\mathrm{eq}}$ is dominant all the time over the relaxation terms
$\left(t^{1-\alpha_{\pm}}\right)$. Namely, in these cases, the transient aging
should not be observed.

Numerical simulations for the cases (2-1) and (2-2) are shown in
Fig.~\ref{f.MSD_case2}(a) and (b), respectively. As predicted, the EMSD and
ETMSD show only normal diffusion. Moreover, the theory and simulations are
consistent without any fitting parameters.

\subsection {MSDs for case (3)}

\begin{figure}[t!]
  \includegraphics[width=56 mm]{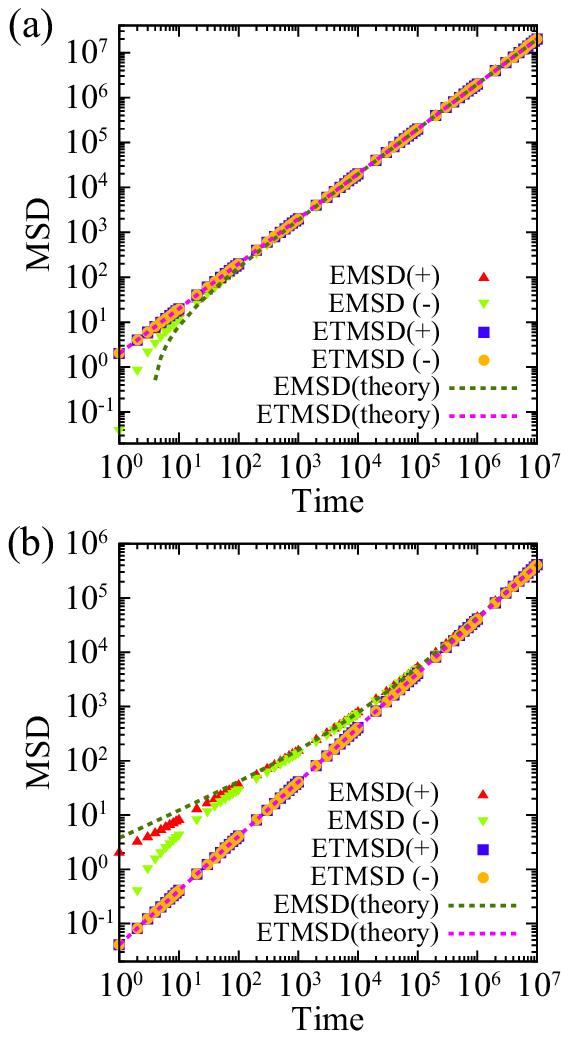}
  \caption{\label{f.MSD_case3}Non-equilibrium EMSD $\left\langle \delta \bm{r}^2
    (\Delta) \right\rangle_{\mathrm{neq}}$ and ETMSD $\langle
    \overline{\delta\bm{r}^2}(\Delta; t) \rangle_{\mathrm{neq}}$ 
    vs lag time $\Delta$ for the case (3).  The total measurement time $t$,
    the initial fraction $p_+^0$, the distance and time units, and the ratio
    $D_+/D_-$ are the same as in Fig.~\ref{f.MSD_case1}.
    (a) [Case (3-1)] $\alpha_{\pm}$ are set as $(\alpha_+, \alpha_-) = (0.5,
    1.5)$. The lines are theoretical predictions given by
    Eqs.~(\ref{e.EMSD(t).case(3a)}) and (\ref{e.ETMSD(t).case(3a)}). (b) [Case
    (3-2)] $\alpha_{\pm}$ are set as $(\alpha_+, \alpha_-) = (1.5, 0.5)$. The
    lines are theoretical predictions given by Eqs.~(\ref{e.EMSD(t).case(3b)})
    and (\ref{e.ETMSD(t).case(3b)}). The crossover time $t_c$ estimated with
    Eq.~(\ref{e.EMSD.case(3b).t_c}) is about $8\times10^2$. }
\end{figure}

Putting Eq.~(\ref{e.p(s).case(3a)}) into Eq.~(\ref{e.EMSD.general.noneq}), we
have the EMSD for the case (3-1) with Laplace inversion:
\begin{equation}
  \label{e.EMSD(t).case(3a)}
  \!
  \left\langle  \delta\bm{r}^2(t) \right\rangle_{\mathrm{neq}} 
  \underset{t \to \infty}{\simeq}
  2n\left[
  D_+ t - \frac {\mu_-}{a_+}
  \frac {(D_+ - D_-) t^{\alpha_+}}{\Gamma (\alpha_+ + 1)}
  \right].
\end{equation}
For this case, from the first inequality in Eq.~(\ref{e.p(t).case(3).range}) and
$D_+ > D_+ - D_-$, we can conclude that the first term in the bracket is
dominant over the second, and therefore the EMSD shows only normal diffusion.

On the other hand, for the case (3-2), we have
\begin{equation}
  \label{e.EMSD(t).case(3b)}
  \!
  \left\langle  \delta\bm{r}^2(t) \right\rangle_{\mathrm{neq}}
  \underset{t \to \infty}{\simeq}
  2n \left[
  D_- t + \frac {\mu_+}{a_-}
  \frac {(D_+ - D_-) t^{\alpha_-}}{\Gamma (\alpha_- + 1)}
  \right].
\end{equation}
In this case, we can observe the transient subdiffusion, if $D_- \ll D_+$. In
fact, the crossover time $t_c$ from subdiffusion to normal diffusion is given by
\begin{equation}
  \label{e.EMSD.case(3b).t_c}
  \frac {\mu_+}{a_-} t_c^{\alpha_- - 1} \sim \frac {D_-}{D_+ - D_-},
\end{equation}
and if this crossover time $t_c$ satisfies the second inequality in
Eq.~(\ref{e.p(t).case(3).range}), i.e., if $D_-/(D_+ - D_-) \ll 1$, we can
observe the transient subdiffusion; note that this condition is equivalent to
$D_- \ll D_+$. In particular, if $D_-=0$, $t_c$ diverges to infinity; thus, the
regime of normal diffusion vanishes, and the EMSD shows only subdiffusion as is
the case for the CTRW. Note that in this case the exponent of the EMSD is the
same as that of CTRW.

\begin{figure*}[]
  \includegraphics[width=160 mm]{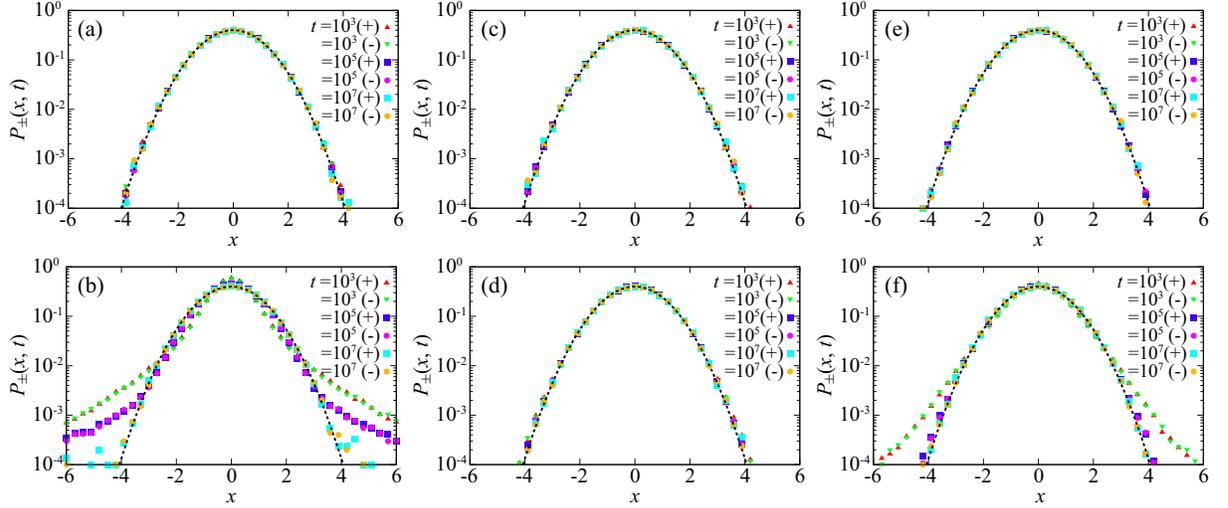}
  \caption{\label{f.propagator} Non-equilibrium propagator
    $P_{\pm}^{\mathrm{neq}}(x, t)$ at three different times $t = 10^3, 10^5$ and
    $10^7$ (each data is normalized so that its variance equals unity).  Two
    non-equilibrium ensembles are used; one is started from $+$ state
    ($p_+^0=1$: triangle up and square), while the other from $-$ state
    ($p_+^0=0$: triangle down and circle). The ratio $D_+/D_-$ is set as
    $D_+/D_- = 50$. The exponents $\alpha_{\pm}$ are set as follows: %
    (a) [case(1-1)] $(\alpha_+, \alpha_-)=(0.25, 0.75)$, %
    (b) [case(1-2)] $(\alpha_+, \alpha_-)=(0.75, 0.25)$, %
    (c) [case(2-1)] $(\alpha_+, \alpha_-)=(1.25, 1.75)$, %
    (d) [case(2-2)] $(\alpha_+, \alpha_-)=(1.75, 1.25)$, %
    (e) [case(3-1)] $(\alpha_+, \alpha_-)=(0.5, 1.5)$, %
    (f) [case(3-2)] $(\alpha_+, \alpha_-)=(1.5, 0.5)$. %
    In (a) and (e), $P_+^{\mathrm{neq}}(x, t)$ is displayed; in (b) and (f),
    $P_-^{\mathrm{neq}}(x, t)$ is displayed; in (c) and (d)
    $P_+^{\mathrm{neq}}(x, t) + P_-^{\mathrm{neq}}(x, t)$ is displayed. The
    dashed lines are the Gaussian distribution with unit variance.}
\end{figure*}


By using Eqs.~(\ref{e.etmsd(t).non-eq}) and (\ref{e.EMSD(t).case(3a)}), we
obtain the ETMSD for the case (3-1) ($\alpha_+ < \alpha_-$) as
\begin{equation}
  \label{e.ETMSD(t).case(3a)}
  \frac
  {\bigl\langle\overline{\delta\bm{r}^2}(\Delta; t)\bigr\rangle_{\mathrm{neq}}}
  {2n\Delta}
  \underset{\begin{subarray}{c}
      \Delta \ll t \\[.03cm] t\to \infty
    \end{subarray}}{\simeq}
  D_+ - \frac {\mu_-}{a_+}
  \frac {(D_+ - D_-) t^{\alpha_+-1}}{\Gamma (\alpha_+ + 1)}.
\end{equation}
For this case, the relaxation of the effective DC is difficult to observe, since
the first term is dominant over the second all the time due to the first
inequality in Eq.~(\ref{e.p(t).case(3).range}) and $D_+ > D_+ - D_-$ (i.e.,
there is no crossover or transient aging).

Similarly, we have the ETMSD for the case (3-2) by using
Eqs.~(\ref{e.etmsd(t).non-eq}) and (\ref{e.EMSD(t).case(3b)}) as follows:
\begin{equation}
  \label{e.ETMSD(t).case(3b)}
  \frac
  {\bigl\langle\overline{\delta\bm{r}^2}(\Delta; t)\bigr\rangle_{\mathrm{neq}}}
  {2n\Delta}
  \underset{\begin{subarray}{c}
      \Delta \ll t \\[.03cm] t\to \infty
    \end{subarray}}{\simeq}
  D_-  + \frac {\mu_+}{a_-}
  \frac {(D_+ - D_-) t^{\alpha_- - 1}}{\Gamma (\alpha_- + 1)}.
\end{equation}
Thus, the effective DC shows a slowing down, and its exponent, $\alpha_- - 1$,
is equivalent to that of CTRW \cite{he08}.  In contrast to the case (3-1), the
relaxation of the effective DC can be observed, because there is a crossover
between relaxation ($t^{\alpha_- - 1}$) at short time to a plateau ($D_-$) at
longer time; the crossover time $t_c$ is given by
Eq.~(\ref{e.EMSD.case(3b).t_c}). That is, the system shows transient aging in
this case.

The theoretical predictions for the EMSD [Eqs.~(\ref{e.EMSD(t).case(3a)}) and
(\ref{e.EMSD(t).case(3b)})] and for the ETMSD [Eqs.~(\ref{e.ETMSD(t).case(3a)})
and (\ref{e.ETMSD(t).case(3b)})] are confirmed by numerical simulations
in Fig.~\ref{f.MSD_case3}. As predicted, for $\alpha_+ < \alpha_-$
[case (3-1)], the EMSD shows normal diffusion [Fig.~\ref{f.MSD_case3}(a)],
whereas, for $\alpha_- < \alpha_+$ [case (3-2)], the EMSD shows transient
subdiffusion [Fig.~\ref{f.MSD_case3}(b)].


\section {Case study: propagator}\label{s.case_study_propagator}

In this section, we derive the non-equilibrium propagator,
$P_{\pm}^{\mathrm{neq}}(\bm{x}, t)$, at the long time limit [The equilibrium
propagator, $P_{\pm}^{\mathrm{eq}}(\bm{x}, t)$, is already derived in
Eq.~(\ref{e.propagator.eq})]. As with the equilibrium process, the propagator
has a non-Gaussian shape at short times, but in a hydrodynamic limit, it
converges to Gaussian distribution if $\alpha_+\neq \alpha_-$. However, this
convergence to Gaussian is very slow for the cases (1-2) and (3-2) [the
slow-mode-dominated cases].

\subsection {Propagator for cases (1) and (3)}

For the case (1--1) ($\alpha_+ < \alpha_-$), using
Eqs.~(\ref{e.rho(s).asymptotic.alpha<1}) and (\ref{e.P(k,s).general.neq}) along
with a hydrodynamic limit $s \sim \bm{k}^2 \ll 1$, we have the leading term of
the propagator
\begin{equation}
  \label{e.P(k,s).case(1-1)}
  \breve{P}_+^{\mathrm{neq}} (\bm{k}, s)
  \underset{
    \begin{subarray}{c}
      s, \bm{k} \to 0 \\[-.03cm]  s \sim \bm{k}^2
    \end{subarray}}{\simeq}
  \frac 1{s + D_+ \bm{k}^2},
\end{equation}
and thus the propagator is Gaussian in this limit. On the other hand,
$\breve{P}_-^{\mathrm{neq}} (\bm{k}, s)$ is negligibly small compared with
$\breve{P}_+^{\mathrm{neq}} (\bm{k}, s)$.  For the case (3--1), exactly the same
results can be obtained. Note also that the propagator given by
Eq.~(\ref{e.P(k,s).case(1-1)}) is that of the diffusion process with a diffusion
constant $D_+$. This is because, for these cases, the fast mode is
asymptotically dominant over the slow mode as can be seen from
Eq.~(\ref{e.p(t).case(1-1)}), and thus long time behavior is not distinguishable
from the single-mode diffusion with the diffusion constant $D_+$
\cite{hofling13}.

Similarly, for the case (1--2) ($\alpha_- < \alpha_+$), we have the following
propagator in the hydrodynamic limit:
\begin{equation}
  \label{e.P(k,s).case(1-2)}
  \breve{P}_-^{\mathrm{neq}} (\bm{k}, s)
  \underset{
    \begin{subarray}{c}
      s, \bm{k} \to 0 \\[-.03cm]  s \sim \bm{k}^2
    \end{subarray}}{\simeq}
  \frac 1{s + D_- \bm{k}^2},
\end{equation}
and $\breve{P}_+^{\mathrm{neq}} (\bm{k}, s)$ is negligibly small compared with
$\breve{P}_-^{\mathrm{neq}} (\bm{k}, s)$.  For the case (3--2), exactly the same
results can be obtained.  As expected, in these cases (1-2) and (3-2), the slow
mode is asymptotically dominant over the fast one.

As shown in Fig.~\ref{f.propagator}, the convergence to the Gaussian
distribution for $\alpha_- < \alpha_+$ [cases (1-2) and (3-2)] is quite slow and
long tails appear [Fig.~\ref{f.propagator}(b) and (f)]. In contrast, for
$\alpha_+ < \alpha_-$ [cases (1-1) and (3-1)] the convergence is relatively fast
[Fig.~\ref{f.propagator}(a) and (e)].

\subsection {Propagator for case (2)}

For the case (2), using Eqs.~(\ref{e.rho(s).asymptotic.alpha>1}) and
(\ref{e.P(k,s).general.neq}) along with the hydrodynamic limit $s \sim \bm{k}^2
\ll 1$, we have the leading term of the propagator
\begin{equation}
  \label{e.P(k,s).case(2)}
  \breve{P}_{\pm}^{\mathrm{neq}} (\bm{k}, s)
    \underset{
      \begin{subarray}{c}
        s, \bm{k} \to 0 \\[-.03cm]  s \sim \bm{k}^2
      \end{subarray}}{\simeq}
  \frac {p_{\pm}^{\mathrm{eq}}}{s + D_{\mathrm{eq}} \bm{k}^2},
\end{equation}
and therefore the propagator for each state, ${P}_{\pm}^{\mathrm{neq}} (\bm{x},
t)$, has a Gaussian shape in the hydrodynamic limit, and the sum
${P}_+^{\mathrm{neq}} (\bm{x}, t) + {P}_-^{\mathrm{neq}} (\bm{x}, t)$ also
becomes a Gaussian distribution. [Note that these propagators are the same as
those of the equilibrium processes given in Eq.~(\ref{e.propagator.eq})]. In
these cases, both fast and slow modes coexist; this is completely different
behavior from the cases (1) and (3).  As shown in Fig.~\ref{f.propagator}(c) and
(d), the convergence to the Gaussian distribution is fast in this case.

\section {Case study: RSD}\label{s.case_study_rsd}

Here, we derive the RSD, $\Sigma(\Delta; t| \bm{w}_0)$, for both equilibrium and
non-equilibrium processes, and show that it exhibits slow relaxation as a
consequence of the fluctuating diffusivity (except for some parameter ranges in
which the RSD shows normal relaxation). As shown in the previous sections, for
the cases (1-2) and (3-2), the EMSD and ETMSD exhibit anomalous behaviors such
as the transient subdiffusion and transient aging. In contrast, for the other
cases including equilibrium processes, both EMSD and ETMSD show normal diffusion
and they do not exhibit (or it is difficult to observe) any anomalous
behaviors. Therefore, for these cases, we can not find any traces of the
fluctuating diffusivity with these MSDs. To find the anomaly of diffusivity, we
have to study higher order moments such as the RSD.


\subsection {Equilibrium RSD}\label{s.case_study_rsd_eq}

\begin{figure}[t!]
  \includegraphics[width=55 mm]{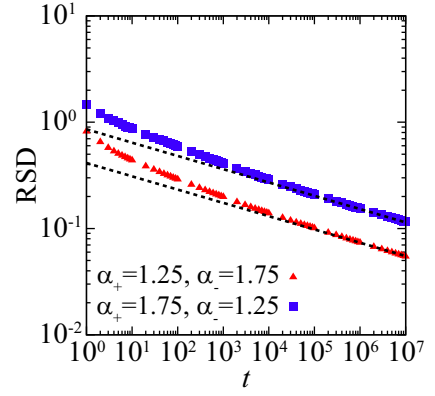}
  \caption{\label{f.rsd.eq} Equilibrium RSD $\Sigma(\Delta; t|
    \bm{w}^{\mathrm{eq}})$ vs total measurement time $t$ for the case (2). The
    exponents $\alpha_{\pm}$ are set as follows: $(\alpha_+, \alpha_-)=(1.25,
    1.75)$ [case(2--1): triangle up] and $(\alpha_+, \alpha_-)=(1.75, 1.25)$
    [case(2--2): square]. The lag time $\Delta$ is set as $\Delta =0.1$. The
    lines are the theoretical prediction [Eqs.~(\ref{e.RSD(t)-case(2-1)-eq}) and
    (\ref{e.RSD(t)-case(2-2)-eq})]. There are no fitting parameters (The same is
    true of Figs.~\ref{f.rsd.neq.case1-case2} and \ref{f.rsd.neq.case3}).  }
\end{figure}


First, we consider the equilibrium ACF $\langle \delta D(t) \delta
D(0) \rangle_{\mathrm{eq}}$ by using Eq.~(\ref{e.<D(s)D(0)>}). We study
only the case (2), since the other cases does not have equilibrium states. With
Eqs.~(\ref{e.rho(s).asymptotic.alpha>1}) and (\ref{e.W(s).eq}), the transition
probabilities for $\alpha_+ < \alpha_-$ [case (2-1)] are found to be
\begin{align}
  \begin{split}
    \hat{W}_{\pm\pm}^{\mathrm{eq}}(s)
    &
    \underset{s \to 0}{\simeq}
    \frac {\left(p_{\pm}^{\mathrm{eq}}\right)^2}{s}
    +
    \frac {a_+\left(p_-^{\mathrm{eq}}\right)^2}{\mu s^{2-\alpha_+}}, 
    \\[0.0cm]
    \hat{W}_{\pm\mp}^{\mathrm{eq}}(s)
    &\underset{s \to 0}{\simeq}
    \frac {p_+^{\mathrm{eq}} p_-^{\mathrm{eq}}}{s}
    -
    \frac {a_+\left(p_-^{\mathrm{eq}}\right)^2}{\mu s^{2-\alpha_+}}. 
  \end{split}
\end{align}
Then, from Eq.~(\ref{e.<D(s)D(0)>}), the ACF is given by
\begin{equation}
  \label{e.<D(s)D(0)>-case(2-1)}
  \left\langle \delta\hat{D}(s) \delta D(0) \right\rangle_{\mathrm{eq}}
  \underset{s \to 0}{\simeq}
  (D_+-D_-)^2 \frac {a_+ (p_-^{\mathrm{eq}})^2}{\mu s^{2-\alpha_+}},
\end{equation}
and the Laplace inversion becomes
\begin{equation}
  \label{e.<D(t)D(0)>-case(2-1)}
  \left\langle \delta D(t) \delta D(0) \right\rangle_{\mathrm{eq}}
  \underset{t \to \infty}{\simeq}
  \frac
  {(D_+-D_-)^2(p_-^{\mathrm{eq}})^2 }
  {\Gamma (2 - \alpha_+)}
  \frac {a_+}{\mu}t^{1-\alpha_+}.
\end{equation}
Thus, in contrast to the MSDs which behave normally, the ACF show slow
relaxation. By using Eqs.~(\ref{e.RSD(s).equilibrium}) and
(\ref{e.<D(s)D(0)>-case(2-1)}), we obtain the RSD
\begin{equation}
  \label{e.RSD(t)-case(2-1)-eq}
  \Sigma^2(\Delta; t | \bm{w}^{\mathrm{eq}})
  \underset{t \to \infty}{\approx}
  \frac {2(D_+-D_-)^2 (p_-^{\mathrm{eq}})^2 }{D_{\mathrm{eq}}^2
    \Gamma(4-\alpha_+)}
  \frac {a_+}{\mu}t^{1-\alpha_+},
\end{equation}
where we used $\Sigma^2(\Delta; t | \bm{w}^{\mathrm{eq}}) \approx
\Sigma^2_{\mathrm{ex}} (t | \bm{w}^{\mathrm{eq}})$, because the ideal RSD,
$\Sigma^2_{\mathrm{id}} (\Delta; t | \bm{w}^{\mathrm{eq}})$, decays faster than
$\Sigma^2_{\mathrm{ex}} (t | \bm{w}^{\mathrm{eq}})$ (see the last
paragraph of Sec.~\ref{s.rsd.general}). The same asymptotic approximation is
used repeatedly in the following subsections.

The ACF shows slow relaxation also for $\alpha_- < \alpha_+$ [case(2-2)] as
\begin{equation}
  \label{e.<D(t)D(0)>-case(2-2)}
  \!
  \left\langle \delta D(t) \delta D(0) \right\rangle_{\mathrm{eq}}
  \underset{t \to \infty}{\simeq}
  \frac
  {(D_+-D_-)^2(p_+^{\mathrm{eq}})^2 }
  {\Gamma (2 - \alpha_-)}
  \frac {a_-}{\mu}t^{1-\alpha_-}.\!
\end{equation}
This equation can be obtained by using Eqs.~(\ref{e.rho(s).asymptotic.alpha>1})
and (\ref{e.W(s).eq}), or more simply, by substituting $\pm$ signs of subscripts
in Eq.~(\ref{e.<D(t)D(0)>-case(2-1)}) for $\mp$ signs. Also, we have the RSD
\begin{equation}
  \label{e.RSD(t)-case(2-2)-eq}
  \Sigma^2(\Delta; t| \bm{w}^{\mathrm{eq}})
  \underset{t \to \infty}{\approx}
  \frac {2(D_+-D_-)^2 (p_+^{\mathrm{eq}})^2}{D_{\mathrm{eq}}^2
    \Gamma(4-\alpha_-)}
  \frac {a_-}{\mu}t^{1-\alpha_-}.
\end{equation}
Thus, in these cases, the RSD, $\Sigma(\Delta; t| \bm{w}^{\mathrm{eq}})$, decays
slower than $t^{-1/2}$; this is the same behavior as the CTRW with the power-law
exponent $\alpha \in (1,2)$ of the waiting time PDF \cite{akimoto11}.

Note also that Eqs.~(\ref{e.RSD(t)-case(2-1)-eq}) and
(\ref{e.RSD(t)-case(2-2)-eq}) are valid for $\Delta$ smaller than the
characteristic correlation time of $D(t)$ [This is the assumption used in the
derivation of Eq.~(\ref{e.RSD(t).general})].
We show results of numerical simulations for the cases (2-1) and (2-2) in
Fig.~\ref{f.rsd.eq}.  These results are found to be consistent with the
theoretical predictions.

\subsection {Non-equilibrium RSD for case (1)}

To derive expressions for the non-equilibrium RSD, we start with the estimate of
$\delta\breve{W}_{hh'}^{\mathrm{neq}} (s_1, s_2)$, because the excess RSD,
$\Sigma_{\mathrm{ex}}(t_1, t_2 | \bm{w}^{\mathrm{neq}})$, is obtained from the
ACF $\langle \delta \hat{D}(s_1) \delta \hat{D}(s_2)\rangle_{\mathrm{neq}}$
[Eq.~(\ref{e.RSD(s).non-equilibrium})], which in turn can be expressed with
$\delta\breve{W}_{hh'}^{\mathrm{neq}} (s_1, s_2)$
[Eq.~(\ref{e.<dD(s1)dD(s2)>})].  For $\alpha_+ < \alpha_-$ [case (1-1)], from
Eqs.~(\ref{e.rho(s).asymptotic.alpha<1}), (\ref{e.def.delta_W(t1,t2)}),
(\ref{e.p(s).case(1-1)}), (\ref{e.app.w(u;s).neq}), (\ref{e.W(s1,s2).neq}), we
obtain
\begin{align}
  \label{e.delta.Whh(s1,s2).case(1-1)}
  \begin{split}
  \!\!\delta \breve{W}_{\pm\pm}^{\mathrm{neq}} (s_1, s_2)\!
  &\underset{
    \begin{subarray}{c}
      s_1, s_2 \to 0 \\[-.0cm]  s_1 \sim s_2
    \end{subarray}}{\simeq}
  \frac {a_-}{a_+}
  \frac {s_1^{\alpha_-}+s_2^{\alpha_-}-(s_1+s_2)^{\alpha_-}}{s_1s_2(s_1+s_2)^{\alpha_+}},\!\!
  \\[.0cm]
  \!\!\delta \breve{W}_{\pm\mp}^{\mathrm{neq}}(s_1, s_2)\!
  &\underset{
    \begin{subarray}{c}
      s_1, s_2 \to 0 \\[-.0cm]  s_1 \sim s_2
    \end{subarray}}{\simeq}
  \frac {a_-}{a_+}
  \frac {(s_1+s_2)^{\alpha_-}-s_1^{\alpha_-}-s_2^{\alpha_-}}{s_1s_2(s_1+s_2)^{\alpha_+}},\!\!
  \end{split}
\end{align}
where we assume that $s_1$ and $s_2$ are comparable ($s_1 \sim s_2$). In
deriving the above equation for $\delta\breve{W}_{-+} (s_1, s_2)$, we used the
relation $\delta\breve{W}_{-+} (s_1, s_2) = \delta\breve{W}_{+-} (s_2, s_1)$,
which is easily checked by Eqs.~(\ref{e.def.delta_W(t1,t2)}),
(\ref{e.app.w(u;s).neq}) and (\ref{e.W(s1,s2).neq}).  From
Eq.~(\ref{e.<dD(s1)dD(s2)>}), we obtain
\begin{align}
  \label{e.<dD(s1)dD(s2)>/s1s2.case(1-1)}
  \frac {\langle \delta \hat{D}(s_1) \delta \hat{D}(s_2)\rangle_{\mathrm{neq}}}{s_1s_2}
  &\underset{
    \begin{subarray}{c}
      s_1, s_2 \to 0 \\[-.03cm]  s_1 \sim s_2
    \end{subarray}}{\simeq}
  \notag\\[.0cm]
  (D_+-D_-)^2&
  \frac {a_- }{a_+}
  \frac {s_1^{\alpha_-}+s_2^{\alpha_-}-(s_1+s_2)^{\alpha_-}}{s_1^2 s_2^2(s_1+s_2)^{\alpha_+}},
\end{align}
Setting $t_1=t_2=t$ in the double Laplace inversions of
Eq.~(\ref{e.<dD(s1)dD(s2)>/s1s2.case(1-1)}) (see Appendix
\ref{s.double.laplace.inversion}), we obtain from
Eqs.~(\ref{e.RSD(s).non-equilibrium}) and (\ref{e.EMSD(t).case(1-1)})
\begin{equation}
  \label{e.RSD(t)-case(1-1)}
  \Sigma^2(\Delta; t| \bm{w}^{\mathrm{neq}})
  \underset{
    \begin{subarray}{c}
      \Delta \ll t \\[.03cm] t\to \infty
    \end{subarray}}{\approx}
  \left(\frac {D_+-D_-}{D_+}\right)^2\frac {a_- }{a_+}
  \frac {2 (1-\alpha_-)\,t^{\alpha_+-\alpha_-}}{\Gamma(3+\alpha_+-\alpha_-)},
\end{equation}
for $\alpha_+<\alpha_-$ [case (1-1)].

On the other hand, we obtain the RSD for $\alpha_-<\alpha_+$ [case (1-2)], by
substituting $\pm$ signs of the subscripts in Eq.~(\ref{e.RSD(t)-case(1-1)}) for
$\mp$ signs,
\begin{equation}
  \label{e.RSD(t)-case(1-2)}
  \Sigma^2(\Delta; t| \bm{w}^{\mathrm{neq}})
  \underset{
    \begin{subarray}{c}
      \Delta \ll t \\[.03cm] t\to \infty
    \end{subarray}}{\approx}
  \left(\frac {D_+-D_-}{D_-}\right)^2\frac {a_+ }{a_-}
  \frac {2(1-\alpha_+)t^{\alpha_--\alpha_+}}{\Gamma(3+\alpha_--\alpha_+)},
\end{equation}
As with the equilibrium RSD, Eqs.~(\ref{e.RSD(t)-case(1-1)}) and
(\ref{e.RSD(t)-case(1-2)}) are valid for $\Delta$ shorter than the
characteristic correlation time of $D(t)$ [The same is true of the results for
the cases (2) and (3) in the following subsections].

\begin{figure}[t!]
  \includegraphics[width=83 mm]{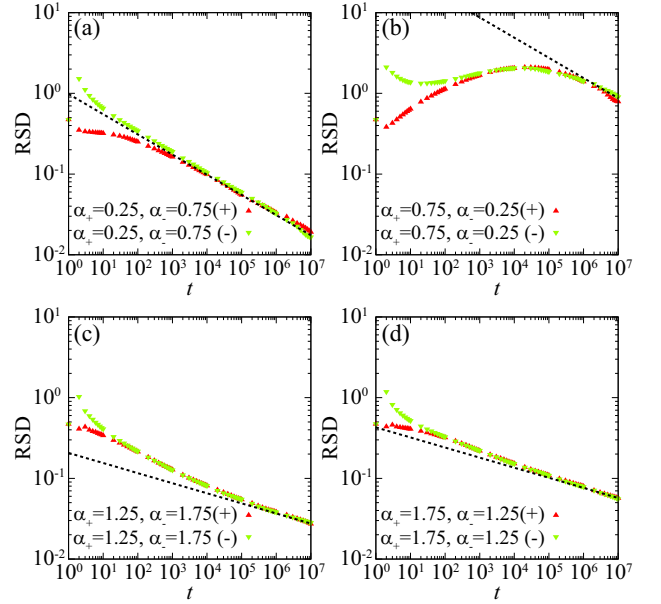}
  \caption{\label{f.rsd.neq.case1-case2}Non-equilibrium RSD $\Sigma(\Delta; t|
    \bm{w}^{\mathrm{neq}})$ vs total measurement time $t$ for the cases (1) and
    (2). $\alpha_{\pm}$ are set as follows: (a) [case(1--1)] $(\alpha_+,
    \alpha_-)=(0.25, 0.75)$, (b) [case(1--2)] $(\alpha_+,
    \alpha_-)=(0.75,0.25)$, (c) [case(2--1)] $(\alpha_+, \alpha_-)=(1.25,
    1.75)$, and (d) [case(2--2)] $(\alpha_+, \alpha_-) =(1.75,1.25)$. For all
    the figures, $\Delta$ is set as $\Delta =0.1$, and results for ensembles
    starting from $+$ state (triangles up) and $-$ state (triangles down) are
    displayed. The lines in (a), (b), (c), and (d) are theoretical predictions
    given in Eqs.~(\ref{e.RSD(t)-case(1-1)}) (\ref{e.RSD(t)-case(1-2)}),
    (\ref{e.RSD(t)-case(2-1)}), and (\ref{e.RSD(t)-case(2-2)}), respectively.}
\end{figure}

In Fig.~\ref{f.rsd.neq.case1-case2} (a) and (b), numerical simulations of the
RSD are displayed for the cases (1-1) and (1-2), respectively. Apparently, the
RSD for the case (1-2) decays slower than that for the case (1-1). This is
consistent with the slow convergence to the Gaussian distribution observed in
Fig.~\ref{f.propagator} (b); note that the excess RSD is equivalent to a
non-Gaussian parameter (see Appendix \ref{s.non-gaussian}).


\subsection {Non-equilibrium RSD for case (2)}

For $\alpha_+ < \alpha_-$ [case (2-1)], from
Eqs.~(\ref{e.rho(s).asymptotic.alpha>1}), (\ref{e.def.delta_W(t1,t2)}),
(\ref{e.p(s).case(2-1)}), (\ref{e.app.w(u;s).neq}), (\ref{e.W(s1,s2).neq}), we
obtain (after a lengthy calculation)
\begin{align}
  \label{e.delta.Whh(s1,s2).case(2-1)}
  \begin{split}
  \delta \breve{W}_{\pm\pm}^{\mathrm{neq}} (s_1, s_2)
  &\underset{
    \begin{subarray}{c}
      s_1, s_2 \to 0 \\[-.0cm]  s_1 \sim s_2
    \end{subarray}}{\simeq}
  \frac {a_+(p_-^{\mathrm{eq}})^2}{\mu}
  \frac {(s_1+s_2)^{\alpha_+}-s_1^{\alpha_+}-s_2^{\alpha_+}}{s_1s_2(s_1+s_2)},
  \\[.0cm]
  \delta \breve{W}_{\pm\mp}^{\mathrm{neq}}(s_1, s_2)
  &\underset{
    \begin{subarray}{c}
      s_1, s_2 \to 0 \\[-.0cm]  s_1 \sim s_2
    \end{subarray}}{\simeq}
  \frac {a_+(p_-^{\mathrm{eq}})^2}{\mu}
  \frac {s_1^{\alpha_+}+s_2^{\alpha_+}-(s_1+s_2)^{\alpha_+}}{s_1s_2(s_1+s_2)},
  \end{split}
\end{align}
where we assume that $s_1$ and $s_2$ are comparable (i.e., $s_1 \sim
s_2$). Thus, from Eq.~(\ref{e.<dD(s1)dD(s2)>}), we obtain
\begin{align}
  \label{e.<dD(s1)dD(s2)>/s1s2.case(2-1)}
  \frac {\langle \delta \hat{D}(s_1) \delta \hat{D}(s_2)\rangle_{\mathrm{neq}}}{s_1s_2}
  &\underset{
    \begin{subarray}{c}
      s_1, s_2 \to 0 \\[-.0cm]  s_1 \sim s_2
    \end{subarray}}{\simeq}
  \\[.0cm]
  \notag
  &\hspace*{-2.0cm}
  \frac {a_+(p_-^{\mathrm{eq}})^2 (D_+-D_-)^2}{\mu}
  \frac {(s_1+s_2)^{\alpha_+} - s_1^{\alpha_+}-s_2^{\alpha_+}}{s_1^2 s_2^2(s_1+s_2)},
\end{align}
Setting $t_1=t_2=t$ in the double Laplace inversions of
Eq.~(\ref{e.<dD(s1)dD(s2)>/s1s2.case(2-1)}),
we obtain from Eqs.~(\ref{e.RSD(s).non-equilibrium}) and
(\ref{e.EMSD(t).case(2-1)})
\begin{equation}
  \label{e.RSD(t)-case(2-1)}
  \Sigma^2(\Delta; t | \bm{w}^{\mathrm{neq}})
  \underset{
    \begin{subarray}{c}
      \Delta \ll t \\[.03cm] t\to \infty
    \end{subarray}}{\approx}
  \frac {2(\alpha_+-1)(D_+-D_-)^2 (p_-^{\mathrm{eq}})^2 }{D_{\mathrm{eq}}^2
    \Gamma(4-\alpha_+)}
  \frac {a_+}{\mu}t^{1-\alpha_+},
\end{equation}
for $\alpha_+<\alpha_-$ [case (2-1)]. On the other hand, by substituting $\pm$
signs of subscripts in Eq.~(\ref{e.RSD(t)-case(2-1)}) for $\mp$ sign, we obtain
the RSD for $\alpha_-<\alpha_+$ [case (2-2)] as
\begin{equation}
  \label{e.RSD(t)-case(2-2)}
  \Sigma^2(\Delta; t| \bm{w}^{\mathrm{neq}})
  \underset{
    \begin{subarray}{c}
      \Delta \ll t \\[.03cm] t\to \infty
    \end{subarray}}{\approx}
  \frac {2(\alpha_--1)(D_+-D_-)^2 (p_+^{\mathrm{eq}})^2 }{D_{\mathrm{eq}}^2
    \Gamma(4-\alpha_-)}
  \frac {a_-}{\mu}t^{1-\alpha_-}.
\end{equation}
Note that the asymptotic scalings $\left(t^{1-\alpha_{\pm}}\right)$ of
Eqs.~(\ref{e.RSD(t)-case(2-1)}) and (\ref{e.RSD(t)-case(2-2)}) are the same as
those of the equilibrium case [Eqs.~(\ref{e.RSD(t)-case(2-1)-eq}) and
(\ref{e.RSD(t)-case(2-2)-eq})]. But there are slight differences in prefactors;
i.e., the non-equilibrium RSDs are smaller than the equilibrium ones.

\subsection {Non-equilibrium RSD for case (3): $\alpha_+ + \alpha_- < 2$}\label{s.rsd.case3.1}

\begin{figure}[t!]
  \includegraphics[width=83 mm]{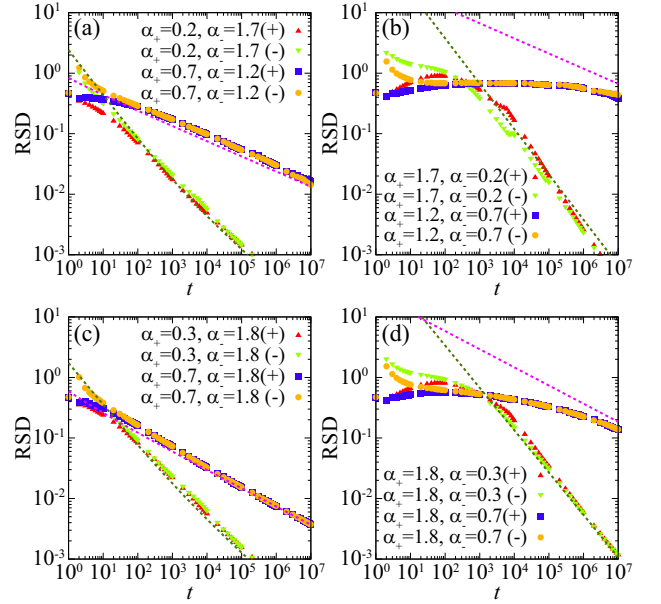}
  \caption{\label{f.rsd.neq.case3}Non-equilibrium RSD $\Sigma(\Delta; t|
    \bm{w}^{\mathrm{neq}})$ vs total measurement time $t$ for the cases (3-1)
    and (3-2). $\alpha_{\pm}$ are set as follows: (a) [case(3-1) with
    $\alpha_++\alpha_- < 2$] $(\alpha_+, \alpha_-)=(0.2, 1.7)$ and $(0.7, 1.2)$,
    (b) [case(3-2) with $\alpha_++\alpha_- < 2$] $(\alpha_+, \alpha_-)=(1.7,
    0.2)$ and $(1.2, 0.7)$ (c) [case(3-1) with $\alpha_++\alpha_- > 2$]
    $(\alpha_+, \alpha_-)=(0.3, 1.8)$ and $(0.7, 1.8)$, and (d) [case(3-2) with
    $\alpha_++\alpha_- > 2$] $(\alpha_+, \alpha_-)=(1.8, 0.3)$ and $(1.8,
    0.7)$. For all the figures, $\Delta$ is set as $\Delta =0.1$, and results
    for ensembles starting from $+$ state (triangles up and squares) and $-$
    state (triangles down and circles) are displayed.  The pink lines in (a),
    (b), (c), and (d) are theoretical predictions for $\Sigma_{\mathrm{ex}}(t|
    \bm{w}^{\mathrm{neq}})$ given by Eqs.~(\ref{e.RSD(t)-case(3a-1)}),
    (\ref{e.RSD(t)-case(3b-1)}), (\ref{e.RSD(t)-case(3a-2)}), and
    (\ref{e.RSD(t)-case(3b-2)}), respectively. Each green line is
    $[4\Delta/(3nt) + \Sigma_{\mathrm{ex}}^2 (t| \bm{w}^{\mathrm{neq}})]^{1/2}$
    [Eq.~(\ref{e.RSD(t)-case(3a-1)-1/t})], in which $\Sigma_{\mathrm{ex}} (t|
    \bm{w}^{\mathrm{neq}})$ is the same as the pink line. }
\end{figure}

For the case (3), the RSD behaves differently depending whether $\alpha_+ +
\alpha_- < 2$ or $\alpha_+ + \alpha_- > 2$. Let us begin with the case of
$\alpha_+ + \alpha_- < 2$.  For the case (3-1) $(\alpha_+ < \alpha_-)$, we
obtain results similar to the case (1)
[Eqs.~(\ref{e.delta.Whh(s1,s2).case(1-1)})--~(\ref{e.RSD(t)-case(1-2)})] as
follows:
\begin{align}
  \label{e.delta.Whh(s1,s2).case(3a-1)}
  \begin{split}
  \!\!\!\delta \breve{W}_{\pm\pm}^{\mathrm{neq}} (s_1, s_2)\!
  &\underset{
    \begin{subarray}{c}
      s_1, s_2 \to 0 \\[-.0cm]  s_1 \sim s_2
    \end{subarray}}{\simeq}
  \frac {a_-}{a_+}
  \frac {(s_1+s_2)^{\alpha_-}-s_1^{\alpha_-}-s_2^{\alpha_-}}{s_1s_2(s_1+s_2)^{\alpha_+}},
  \!\\[.0cm]
  \!\!\!\delta \breve{W}_{\pm\mp}^{\mathrm{neq}}(s_1, s_2)\!
  &\underset{
    \begin{subarray}{c}
      s_1, s_2 \to 0 \\[-.0cm]  s_1 \sim s_2
    \end{subarray}}{\simeq}
  \frac {a_-}{a_+}
  \frac {s_1^{\alpha_-}+s_2^{\alpha_-}-(s_1+s_2)^{\alpha_-}}{s_1s_2(s_1+s_2)^{\alpha_+}},
  \!
  \end{split}
\end{align}
where we used Eqs.~(\ref{e.rho(s).asymptotic.alpha<1}),
(\ref{e.rho(s).asymptotic.alpha>1}), (\ref{e.def.delta_W(t1,t2)}),
(\ref{e.p(s).case(3a)}), (\ref{e.p(s).case(3a)-higher-order-terms}),
(\ref{e.app.w(u;s).neq}), and (\ref{e.W(s1,s2).neq}) \cite{Note2}. These
equations are almost the same as Eq.~(\ref{e.delta.Whh(s1,s2).case(1-1)}) except
for their signs. Thus, the RSD is obtained by changing the sign of
Eq.~(\ref{e.RSD(t)-case(1-1)}) as
\begin{equation}
  \label{e.RSD(t)-case(3a-1)}
  \Sigma^2_{\mathrm{ex}}(t| \bm{w}^{\mathrm{neq}})
  \underset{
    \begin{subarray}{c}
      t\to \infty
    \end{subarray}}{\simeq}
  \left(\frac {D_+-D_-}{D_+}\right)^2\frac {a_- }{a_+}
  \frac {2 (\alpha_- - 1)\,t^{\alpha_+-\alpha_-}}{\Gamma(3+\alpha_+-\alpha_-)},
\end{equation}
for $\alpha_+<\alpha_-$ [case(3-1)].  Note that $\Sigma^2_{\mathrm{ex}}(t|
\bm{w}^{\mathrm{neq}})$ decays faster than $O(1/t)$, if $\alpha_- - \alpha_+ >
1$. In such a case, the ideal RSD, $\Sigma^2_{\mathrm{id}}(\Delta; t|
\bm{w}^{\mathrm{neq}})$ [Eq.~(\ref{e.RSD(t).sigma_id.asympt.case(3)})], which is
of the order of $O(1/t)$, becomes the dominant term, and the excess RSD,
$\Sigma^2_{\mathrm{ex}}(t| \bm{w}^{\mathrm{neq}})$
[Eqs.~(\ref{e.RSD(t)-case(3a-1)})], becomes the second leading term:
\begin{equation}
  \label{e.RSD(t)-case(3a-1)-1/t}
  \Sigma^2(\Delta; t| \bm{w}^{\mathrm{neq}})
  \underset{
    \begin{subarray}{c}
      \Delta \ll t \\[.03cm] t\to \infty
    \end{subarray}}{\approx}
  \frac {4\Delta}{3nt}
  +
  \Sigma^2_{\mathrm{ex}}(t| \bm{w}^{\mathrm{neq}}).
\end{equation}
In contrast, if $\alpha_- - \alpha_+ < 1$, $\Sigma^2_{\mathrm{ex}}(t|
\bm{w}^{\mathrm{neq}})$ decays slower than $O(1/t)$, and thus this is the
dominant term:
\begin{equation}
  \label{e.RSD(t)-case(3a-1)-slower}
  \Sigma^2(\Delta; t| \bm{w}^{\mathrm{neq}})
  \underset{
    \begin{subarray}{c}
      \Delta \ll t \\[.03cm] t\to \infty
    \end{subarray}}{\approx}
  \Sigma^2_{\mathrm{ex}}(t| \bm{w}^{\mathrm{neq}}).
\end{equation}
As shown in Fig.~\ref{f.rsd.neq.case3}(a), results of numerical simulations are
consistent with these asymptotic behaviors.


On the other hand, we obtain the RSD for the case (3-2) $(\alpha_-<\alpha_+)$ by
replacing $\pm$ signs of subscripts in Eq.~(\ref{e.RSD(t)-case(3a-1)}) with
$\mp$ signs:
\begin{equation}
  \label{e.RSD(t)-case(3b-1)}
  \Sigma^2_{\mathrm{ex}}(t| \bm{w}^{\mathrm{neq}})
  \underset{
    \begin{subarray}{c}
      t\to \infty
    \end{subarray}}{\simeq}
  \left(\frac {D_+-D_-}{D_-}\right)^2\frac {a_+ }{a_-}
  \frac {2(\alpha_+ - 1)\,t^{\alpha_--\alpha_+}}{\Gamma(3+\alpha_--\alpha_+)}.
\end{equation}
As with the case (3-1), $\Sigma^2_{\mathrm{ex}}(t| \bm{w}^{\mathrm{neq}})$
decays faster than $O(1/t)$, if $\alpha_+ - \alpha_- > 1$. Then, the dominant
term is given by $\Sigma^2_{\mathrm{id}}(\Delta; t| \bm{w}^{\mathrm{neq}})$ in
Eq.~(\ref{e.RSD(t).sigma_id.asympt.case(3)}), and thus the asymptotic RSD is
again reduced to Eq.~(\ref{e.RSD(t)-case(3a-1)-1/t}) in which
$\Sigma^2_{\mathrm{ex}}( t| \bm{w}^{\mathrm{neq}})$ is given by
Eq.~(\ref{e.RSD(t)-case(3b-1)}). Contrastingly, if $\alpha_+ - \alpha_- < 1$,
$\Sigma^2_{\mathrm{ex}}(t| \bm{w}^{\mathrm{neq}})$ decays slower than $O(1/t)$,
and thus this term is dominant. Accordingly, the RSD behaves as
Eq.~(\ref{e.RSD(t)-case(3a-1)-slower}) in which $\Sigma^2_{\mathrm{ex}}(t|
\bm{w}^{\mathrm{neq}})$ is given by Eq.~(\ref{e.RSD(t)-case(3b-1)}).
As shown in Fig.~\ref{f.rsd.neq.case3} (b), we found good agreements between
these theories and numerical results. In particular, when $\alpha_+ - \alpha_- <
1$, the RSD decays quite slowly.

\subsection {Non-equilibrium RSD for case (3): $\alpha_+ + \alpha_- > 2$}\label{s.rsd.case3.2}

Now we study the case in which $\alpha_+ + \alpha_- > 2$. Let us start with the
case (3-1) $(\alpha_+<\alpha_-)$.  We obtain the following expressions for
$\delta \breve{W}_{\pm\pm}^{\mathrm{neq}} (s_1, s_2)$ and $\delta
\breve{W}_{\pm\mp}^{\mathrm{neq}} (s_1, s_2)$ from
Eqs.~(\ref{e.rho(s).asymptotic.alpha<1}), (\ref{e.rho(s).asymptotic.alpha>1}),
(\ref{e.def.delta_W(t1,t2)}), (\ref{e.p(s).case(3a)}),
(\ref{e.p(s).case(3a)-higher-order-terms}), (\ref{e.app.w(u;s).neq}), and
(\ref{e.W(s1,s2).neq}) \cite{Note2}:
\begin{align}
  \label{e.delta.Whh(s1,s2).case(3a-2)}
  \begin{split}
    \delta \breve{W}_{\pm\pm}^{\mathrm{neq}} (s_1, s_2)
    &\underset{
      \begin{subarray}{c}
        s_1, s_2 \to 0 \\[-.0cm]  s_1 \sim s_2
      \end{subarray}}{\simeq}
    \left(\frac {\mu_-}{a_+}\right)^2
    \frac {s_1^{\alpha_+}+s_2^{\alpha_+} - (s_1+s_2)^{\alpha_+}}
    {[s_1s_2(s_1+s_2)]^{\alpha_+}},
    \\[.0cm]
    \delta \breve{W}_{\pm\mp}^{\mathrm{neq}}(s_1, s_2)
    &\underset{
      \begin{subarray}{c}
        s_1, s_2 \to 0 \\[-.0cm]  s_1 \sim s_2
      \end{subarray}}{\simeq}
    \left(\frac {\mu_-}{a_+}\right)^2
    \frac {(s_1+s_2)^{\alpha_+} - s_1^{\alpha_+} - s_2^{\alpha_+} }
    {[s_1s_2(s_1+s_2)]^{\alpha_+}},
  \end{split}
\end{align}
where we assume that $s_1$ and $s_2$ are comparable (i.e., $s_1 \sim
s_2$). Thus, from Eq.~(\ref{e.<dD(s1)dD(s2)>}), we obtain
\begin{align}
  \label{e.<dD(s1)dD(s2)>/s1s2.case(3a-2)}
  \frac {\langle \delta \hat{D}(s_1) \delta \hat{D}(s_2)\rangle_{\mathrm{neq}}}{s_1s_2}
  &\underset{\begin{subarray}{c}
    s_1, s_2 \to 0 \\[-.0cm]  s_1 \sim s_2
  \end{subarray}}{\simeq}
(D_+-D_-)^2  \left(\frac {\mu_-}{a_+}\right)^2
  \notag\\[.0cm]
  &\times
  \frac {s_1^{\alpha_+}+s_2^{\alpha_+}-(s_1+s_2)^{\alpha_+}}
  {(s_1s_2)^{\alpha_++1}(s_1+s_2)^{\alpha_+}}.
\end{align}
Setting $t_1=t_2=t$ in the double Laplace inversions of this equation, we obtain
from Eqs.~(\ref{e.RSD(s).non-equilibrium}) and (\ref{e.EMSD(t).case(3a)})
\begin{align}
  \Sigma^2_{\mathrm{ex}}(t| \bm{w}^{\mathrm{neq}})
  &\underset{
    \begin{subarray}{c}
      t\to \infty
    \end{subarray}}{\simeq}
  \left(\frac {D_+-D_-}{D_+}\right)^2
  \left(\frac {\mu_- }{a_+}\right)^2\notag\\[0.0cm]
  \times& \frac {t^{2\alpha_+ - 2}}{\Gamma^2(\alpha_++1)}
  \left[
  \frac {2\Gamma^2(\alpha_++1)}{\Gamma(2\alpha_++1)} - 1
  \right],
  \label{e.RSD(t)-case(3a-2)}
\end{align}
for $\alpha_+<\alpha_-$ [case (3-1)]. Note that, if $\alpha_+ < 1/2$,
$\Sigma^2_{\mathrm{ex}}(t| \bm{w}^{\mathrm{neq}})$ decays faster than
$O(1/t)$. Therefore, the ideal RSD $\Sigma^2_{\mathrm{id}}(\Delta; t|
\bm{w}^{\mathrm{neq}})$ given by Eq.~(\ref{e.RSD(t).sigma_id.asympt.case(3)})
becomes the dominant term, and the excess part $\Sigma^2_{\mathrm{ex}}(
t| \bm{w}^{\mathrm{neq}})$ [Eq.~(\ref{e.RSD(t)-case(3a-2)})] is the second
leading contribution. In this case (i.e., $\alpha_+ < 1/2$), the RSD
asymptotically behaves as Eq.~(\ref{e.RSD(t)-case(3a-1)-1/t}) in which
$\Sigma^2_{\mathrm{ex}}(t| \bm{w}^{\mathrm{neq}})$ is given by
Eq.~(\ref{e.RSD(t)-case(3a-2)}).

\begin{figure}[t!]
  \centerline{\includegraphics[width=7.77cm]{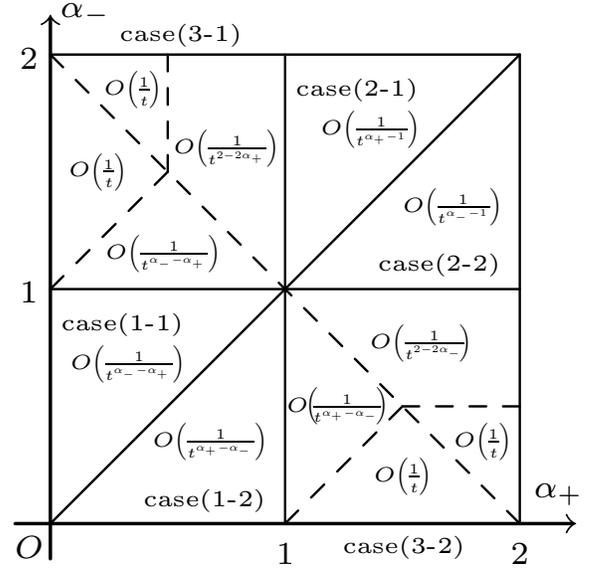}}
  \caption{\label{f.phase-diagram} Phase diagram for asymptotic behaviors of the
    non-equilibrium RSD, $\Sigma^2(\Delta; t| \bm{w}^{\mathrm{neq}})$, on the
    $\alpha_+\alpha_-$ plane. The solid lines represent boundaries dividing the
    six cases (1-1), (1-2), (2-1), (2-2), (3-1), and (3-2). The dashed lines are
    boundaries dividing the cases (3-1) and (3-2) into subclasses according to
    the scaling behavior of the RSD.}
\end{figure}

Moreover, the RSD for the case (3-2) $\left(\alpha_-<\alpha_+\right)$ is
obtained by replacing $\pm$ signs of the subscripts in
Eq.~(\ref{e.RSD(t)-case(3a-2)}) with $\mp$ signs:
\begin{align}
  \Sigma^2_{\mathrm{ex}}(t| \bm{w}^{\mathrm{neq}})
  &\underset{
    \begin{subarray}{c}
      t\to \infty
    \end{subarray}}{\simeq}
  \left(\frac {D_+-D_-}{D_-}\right)^2
  \left(\frac {\mu_+ }{a_-}\right)^2\notag\\[0.0cm]
  \times& 
  \frac {t^{2\alpha_- - 2}}{\Gamma^2(\alpha_-+1)}
  \left[
  \frac {2\Gamma^2(\alpha_-+1)}{\Gamma(2\alpha_-+1)} - 1
  \right].
  \label{e.RSD(t)-case(3b-2)}
\end{align}
Again, the leading term is switched at $\alpha_- = 1/2$ from
$\Sigma^2_{\mathrm{id}}(\Delta; t| \bm{w}^{\mathrm{neq}})$ to
$\Sigma^2_{\mathrm{ex}}(t| \bm{w}^{\mathrm{neq}})$.
As shown in Fig.~\ref{f.rsd.neq.case3}(c) and (d), we checked these asymptotic
behaviors by numerical simulations, and found good agreements with these theory.
The asymptotic scaling behavior of the non-equilibrium RSD is summarized in
Fig.~\ref{f.phase-diagram}, which shows that the RSD exhibits slow relaxation
$O(1/t^{\beta})$ with $\beta < 1$ for most of the parameter values $(\alpha_+,
\alpha_-)$; the exceptions are some parameter domains inside of the cases (3-1)
and (3-2), in which the RSD decays as $O(1/t)$.


\section {Discussion}\label{s.discussion}
\subsection {Summary}

In many biological experiments \cite{golding06,burov11,jeon11, weigel11} and
molecular dynamics simulations \cite{akimoto11, yamamoto14, yamamoto15}, scatter
of DCs has been reported. To model such scatter, CTRW and related models
have been studied extensively, and it is found that these models exhibit
distributional ergodicity---namely, the DC behaves as a random variable
\cite{he08, neusius09, meroz10, miyaguchi11c, miyaguchi13}. This phenomenon is
reminiscent of the scatter of DC observed in experiments. Some experimental
studies, however, show that the DC temporally fluctuates between two distinct
values \cite{yamamoto98a, yamamoto98b, knight09, leith12, parry14}. The CTRW
type models should not be suitable as a model of such dichotomous fluctuations
of the DC, since these models do not consist of two diffusion modes, but of long
quiescent states and instantaneous jumps.


In this paper, we studied a Langevin equation with two diffusion modes (fast and
slow diffusion modes) for equilibrium and non-equilibrium ensembles.  We found
that the EMSD shows normal diffusion or transient subdiffusion depending on the
sojourn time PDFs, and that the TMSD shows only normal diffusion.  Thus, it is
impossible with the TMSD to find a trace of fluctuating diffusivity. This is a
serious problem, since, in many experiments, the TMSD is frequently utilized to
calculate the MSD. As an alternative approach, we have proposed the RSD analysis
with which we can extract information of the ACF, $\left\langle D(t_1)D(t_2)
\right\rangle$, from trajectory data $\bm{x}(t)$ [namely, the instantaneous DC,
$D(t)$, is not necessary to measure directly]. It is also worth noting that the
excess RSD is equivalent to a non-Gaussian parameter. From these facts, the RSD
analysis should be a very useful tool to analyze the fluctuating diffusivity in
single-particle-tracking experiments.  In the case studies for the power-law
sojourn time PDFs, we also show that the RSD decays slower than $t^{-1/2}$.
Such slow relaxation in RSD is observed in molecular dynamics simulations for
the center-of-mass motion of a lipid molecule in a lipid bilayer
\cite{akimoto11}.

We also show that the propagator has a non-Gaussian shape for short time, and
converges to the Gaussian distribution asymptotically. This convergence to
Gaussian is extremely slow for the cases in which the slow mode dominates over
the fast one [cases (1-2) and (3-2)].
Furthermore, in these slow-mode-dominated cases, the EMSD shows transient
subdiffusion, and the effective DC of the ETMSD exhibits transient aging.
On the other hand, the equilibrium processes show the short-time
non-Gaussianity, and yet normal diffusion in both EMSD and ETMSD. This is
consistent with the findings reported in \cite{mykyta14}, in which a random walk
model with fluctuating diffusivity is studied.

In addition, if we take a limit $D_- \to 0$ in the slow-mode-dominated cases
[cases (1-2) and (3-2)], the behaviors of the system become very similar to
those of CTRW. In fact, the crossover time from the transient to asymptotic
regimes, $t_c$ [Eqs.~(\ref{e.EMSD.case(1-2).t_c}) and
(\ref{e.EMSD.case(3b).t_c})], diverges when $D_- \to 0$, thereby permanent
subdiffusion and aging are observed.
This implies that the scatter of the DC reported in many
single-particle-tracking experiments \cite{golding06,burov11,jeon11} and
molecular dynamics simulations \cite{akimoto11, yamamoto14, yamamoto15}, which
so far has been attributed to CTRW dynamics, may well be originated from the
fluctuating diffusivity.
Moreover, although we studied the LEFD with dichotomous DCs, some of the general
results presented in Sec.~\ref{s.general_theory} are valid for a general
LEFD. In fact, Eqs.~(\ref{e.emsd}), (\ref{e.etmsd(t).final}) and
(\ref{e.RSD(t).general}) are valid for arbitrary stochastic processes $D(t)$.

Also, we confirmed our theoretical results by extensive numerical
simulations. Importantly, we used an experimentally plausible ratio between the
two DCs, $D_+/D_- = 50$ (see \cite{parry14}, in which diffusion in the bacterial
cells is studied. See also Appendix \ref{sec:app.simulation}), and observed
singular behaviors such as the transient subdiffusion, non-Gaussianity of the
propagator, and slow relaxation in the RSD. This means that these singular
behaviors might be measurable also in single-particle-tracking experiments of
bacterial cells.

\subsection {Outlook}

Mechanisms of fluctuating diffusivity observed in crowded and glass-like systems
are still unclear \cite{yamamoto98a,yamamoto98b, parry14} (except for the
reptation dynamics in which fluctuating diffusivity originates from the the
end-to-end vector fluctuations \cite{uneyama15}). Usually, tagged-particle
diffusion can be described by a generalized Langevin equation (GLE). When
deriving the GLE with the projection operator method \cite{mori65}, however, we
have to assume a separation of time scales of the tagged particle (slow
variables) and environment (fast variables). Thus, the fluctuating diffusivity
might be due to a violation of this separation of the time
scales.

It is also worth mentioning the GLE and fractional Brownian motion (FBM)
\cite{deng09, ferreira12}.
The main difference between these systems and the LEFD is in that the GLE and
FBM have Gaussian propagators since they are Gaussian processes, whereas the
LEFD shows non-Gaussian behavior (Though, some recent works studied GLEs with
the L\'evy stable noise, which shows non-Gaussianity \cite{srokowski13}).
Moreover, for the GLE and FBM, the velocity ACF exhibits power law decay
\cite{pottier03}, while that of the LEFD is delta-correlated
\begin{equation}
  \left\langle \bm{v}(t) \bm{v}(t') \right\rangle
  =
  2n \left\langle D(t) \right\rangle \delta(t-t').
\end{equation}
In this sense, both GLE/FMB and LEFD are extreme models, and real systems might
show both properties: non-Gaussianity and slow decay of the velocity ACF.

As presented in \cite{parry14}, relatively large particles in bacterial cells
with a normal metabolism show fast and slow diffusion modes, which motivates us
to study the LEFD with the dichotomous DCs.  On the other hand, in metabolically
dormant cells, this dichotomy of the DC almost vanishes and a single peak at
zero-diffusivity appears in the PDF of the diffusivity (Fig.5B in
\cite{parry14}). To describe this type of behavior, other models such as the
ATTM \cite{massignan14, manzo15} would be important.
Moreover, the TMSD measured in \cite{parry14} shows subdiffusion, while the TMSD
of the LEFD only shows normal diffusion. This discrepancy would be explained by
considering a finite size (confinement) effect \cite{he08, neusius09,
  miyaguchi13}.
  
In this article, we only studied some basic properties of the LEFD such as the
MSDs and propagators, but more advanced properties such as the first passage
time, escape statistics, diffusion on potentials landscapes are also interesting
subjects to study. In particular, the first passage time is important in the
analysis of intermittent search problems \cite{loverdo09,benichou11}. In earlier
studies \cite{reingruber09,reingruber10}, the Markovian switching between fast
and slow diffusion modes is investigated as an efficient search strategy, but
effects of non-Markovian switchings on the search efficiency may be important. To
understand such non-Markovian effects, the two-state renewal theory developed in
this article is essential. This two-state renewal theory will also be utilized
in various fields of science, since many natural systems behave like two-state
systems (e.g., ion-channel gating \cite{goychuk03}, blinking quantum dots
\cite{kuno00, brokmann03}, and Kardar-Parisi-Zhang fluctuations
\cite{takeuchi15}).

In addition, we concentrated on the equilibrium and a typical non-equilibrium
ensembles, and did not analyze aging processes which is assumed to start at
$t=-t_a$ \cite{barkai03a, schulz14}. However, the aging phenomena can be
analyzed to some extent by using the general transition probabilities given in
Eq.~(\ref{e.W(s1,s2).neq}), or more simply by replacing the initial ensemble
$\bm{w}^0(\tau) = \{w^0_{\pm}(\tau)\}$ with the the forward recurrence time PDF
$\bm{w}_a^{\mathrm{neq}}(\tau) := \{w_{\pm}^{\mathrm{neq}} (\tau; t_a)\}$
[Eq.~(\ref{e.app.w(u;s).neq.0})] in various results. For example, $p_{\pm} (t|
\bm{w}_a^{\mathrm{neq}})$ is the fractions at $t$ for a process that started at
$t=-t_a$ with the initial ensemble $\bm{w}^{\mathrm{neq}}(\tau)$.


\begin{acknowledgments}
  TM was supported by JSPS KAKENHI for Young Scientists (B) (Grant No.
  15K17590), and TA was supported by JSPS KAKENHI for Young Scientists (B)
  (Grant No. 22740262).
\end{acknowledgments}

\appendix {}
\section {Another derivation of $\hat{Q}_{\pm,n}(s)$}\label{s.app.Q}

In Sec.~\ref{s.general_theory}, we derive $\hat{Q}_{\pm,n}(s | \bm{w}^0)$
with a straightforward calculation [see Eq.~(\ref{e.Q(s)})]. It is easier to
derive it in the following way. First, let us write $Q_{\pm,n}(t | \bm{w}^0)$ as
\begin{equation}
  \label{e.app.Q(t)}
  Q_{\pm,n}(t | \bm{w}^0)
  =
  \left\langle I(t_n < t < t_{n+1})\, d_{\pm}\right\rangle,
\end{equation}
where $I(\dots) = 1$ if the inside of the bracket is satisfied, 0
otherwise. Moreover, $d_{\pm} := \delta_{D(0), D_{\pm}}$ is a random variable
indicating the initial state, i.e.,
\begin{equation}
  d_{\pm} =
  \begin{cases}
    1 \quad\, & \text{if the state is $\pm$ at $t=0$},\\
    0 \quad\, & \text{if the state is $\mp$ at $t=0$},
  \end{cases}
\end{equation}
and $t_n\,(n=1, 2, \dots)$ are called renewal times, at which the transitions
from one state to the other occur (we also set $t_0 = 0$ for convenience). The
renewal time $t_n\,(n=0, 1, \dots)$ is written by a sum of successive sojourn
times between transitions, $\tau_k$ ($k=1,2,\dots$), as $t_n \equiv
\sum_{k=1}^{n} \tau_k$.

The Laplace transform of Eq.~(\ref{e.app.Q(t)}) gives
\begin{equation}
  \label{e.app.Q(s)}
  \hat{Q}_{\pm,n}(s | \bm{w}^0)
  =
  \begin{cases}
    \displaystyle
    \frac {\left\langle (1 - e^{-s\tau_1})d_{\pm} \right\rangle}{s}
    & (n=0),
    \\[0.0cm]
    \displaystyle
    \left\langle e^{-st_n} d_{\pm}\right\rangle
    \frac {1 - \left\langle e^{-s\tau_{n+1}} \right\rangle}{s}
    & (n=1,2,\dots).
  \end{cases}
\end{equation}
Moreover, we have
\begin{equation}
  \left\langle (1 - e^{-s\tau_1})d_{\pm} \right\rangle
  =
  p_{\pm}^{0} - \hat{w}^0_{\pm}(s),
\end{equation}
because the first sojourn time $\tau_1$ follows the PDF $w_+^0(\tau) +
w_-^0(\tau) = p_+^0\rho_+^0(\tau) + p_-^0\rho_-^0(\tau)$. Similarly, for the
even terms $\hat{Q}_{\pm,2n}(s | \bm{w}^0)$, we have
\begin{align}
  \label{e.app.Q(s).2n.parts1}
  \left\langle e^{-st_{2n}} d_{\pm}\right\rangle
  &=
  \left\langle e^{-s\tau_1} d_{\pm}\right\rangle
  \left\langle e^{-s\tau_2} \right\rangle
  \times\dots\times
  \left\langle e^{-s\tau_{2n}} \right\rangle
  \notag\\[0.0cm]
  &=
  \hat{w}^0_{\pm}(s) \hat{\rho}^{n-1}(s) \hat{\rho}_{\mp}(s),
\end{align}
where $\hat{\rho} (s) = \hat{\rho}_+(s) \hat{\rho}_-(s)$, and 
\begin{equation}
  \label{e.app.Q(s).2n.parts2}
  \frac {1 - \left\langle e^{-s\tau_{2n+1}} \right\rangle}{s}
  =
  \frac {1- \hat{\rho}_{\pm}(s)}{s}.
\end{equation}
Therefore, from Eqs.~(\ref{e.app.Q(s)}), (\ref{e.app.Q(s).2n.parts1}), and
(\ref{e.app.Q(s).2n.parts2}), we obtain Eq.~(\ref{e.Q(s)}) for the even
terms. In the same way, for the odd terms $\hat{Q}_{\pm,2n-1}(s |
\bm{w}^0)$, we obtain
\begin{align}
  \label{e.app.Q(s).2n-1.parts1}
  \left\langle e^{-st_{2n-1}} d_{\pm}\right\rangle
  &=
  \hat{w}^0_{\pm}(s) \hat{\rho}^{n-1}(s),
  \\[0.0cm]
  \label{e.app.Q(s).2n-1.parts2}
  \frac {1 - \left\langle e^{-s\tau_{2n}} \right\rangle}{s}
  &=
  \frac {1- \hat{\rho}_{\mp}(s)}{s}.
\end{align}
Therefore, from Eqs.~(\ref{e.app.Q(s)}), (\ref{e.app.Q(s).2n-1.parts1}), and
(\ref{e.app.Q(s).2n-1.parts2}), we obtain Eq.~(\ref{e.Q(s)}) for the odd terms.

\section {PDF for forward recurrence time}\label{s.pdf-elapsed-time}

A forward recurrence time is a sojourn time until the next transition from some
elapsed time $t' > 0$ \cite{godrche01}.  Here, using a technique in the previous
section, we study the forward recurrence time PDF, $w_\pm(\tau;t'| \bm{w}^0)$.
More precisely, $w_\pm(\tau;t'| \bm{w}^0) d\tau$ is a joint probability that %
(1) the state is $\pm$ at time $t'$, and %
(2) the sojourn time from $t'$ until the next transition is in an interval
$[\tau, \tau+d\tau]$ ($\tau$ is the forward recurrence time), given that the
process starts with $\bm{w}^0$ at $t=0$.
First, let us define
\begin{equation}
  w_{\pm,n}(\tau; t' | \bm{w}^0) :=
  \left\langle
  \delta \left(\tau - (t_{n+1} - t')\right)
  I \left( t_n < t' < t_{n+1} \right)
  d_{\pm}
  \right\rangle,
\end{equation}
where $w_{\pm, n}(\tau; t')d\tau$ is a joint probability that %
(1) the state is $\pm$ at $t=0$, %
(2) there are $n$ transitions until time $t'$, and %
(3) the sojourn time at time $t'$ is in an interval $[\tau, \tau+d\tau]$, %
given that the process starts with $\bm{w}^0$ at $t=0$.
Then, the double Laplace transforms with respect to $\tau$ and $t'$ result in
\begin{align}
  \breve{w}_{\pm,n} (u; s| \bm{w}^0) &=
  \left\langle
  \frac {d_{\pm} e^{-t_ns}}{s-u}
  \left(
  e^{-u\tau_{n+1}}-e^{-s\tau_{n+1}}
  \right)
  \right\rangle
  \\[0.0cm]
  \notag
  &\hspace*{-1.7cm}=
  \begin{cases}
    \displaystyle
    \frac {\left\langle
      d_{\pm}
      \left(
      e^{-u\tau_1}-e^{-s\tau_1}
      \right)
      \right\rangle}{s-u}
    & (n=0),\\[0.20cm]
    \displaystyle
    \left\langle d_{\pm} e^{-st_n} \right\rangle
    \frac {\left\langle
    \left( e^{-u\tau_{n+1}}-e^{-s\tau_{n+1}} \right)
    \right\rangle}{s-u}
    & (n=1,2,\dots).
  \end{cases}
\end{align}
Then, in the same way as the previous section, we have
\begin{align}
  \breve{w}_{\pm,0} (u; s | \bm{w}^0) &=
  \dfrac {\hat{w}_{\pm}^0 (u) - \hat{w}_{\pm}^{0} (s)}{s-u},
  \\[0.0cm]
  \breve{w}_{\pm,2n} (u; s | \bm{w}^0) &=
  \hat{w}^0_{\pm}(s)\hat{\rho}^{n-1} (s) \hat{\rho}_{\mp}(s)
  \dfrac {\hat{\rho}_{\pm} (u) - \hat{\rho}_{\pm} (s)} {s-u},
  \\[0.0cm]
  \breve{w}_{\pm,2n-1} (u; s | \bm{w}^0) &=
  \hat{w}^0_{\pm}(s)\hat{\rho}^{n-1} (s)
  \dfrac {\hat{\rho}_{\mp} (u) - \hat{\rho}_{\mp} (s)} {s-u},
\end{align}
for $n=1,2,\dots$.

Because the Laplace transform of the forward recurrence time PDF, $w_\pm(\tau;t'
| \bm{w}^0)$, is given by $\breve{w}_{\pm} (u; s | \bm{w}^0) =
\sum_{n=0}^{\infty}[\breve{w}_{\pm,2n} (u; s | \bm{w}^0 ) + \breve{w}_{\mp,2n+1}
(u; s | \bm{w}^0)]$, we have
\begin{align}
  \label{e.app.w(u;s).general}
  \breve{w}_{\pm}(u; s | \bm{w}^0) &=
  \frac {\hat{w}^0_{\pm} (u) - \hat{w}^0_{\pm} (s)}{s-u}\notag\\[0.0cm]
  +&
  \frac {\hat{w}^0_{\pm}(s) \hat{\rho}_{\mp}(s) + \hat{w}^0_{\mp}(s)}{1 - \hat{\rho} (s)}
  \frac {\hat{\rho}_{\pm} (u) - \hat{\rho}_{\pm} (s)} {s-u}.
\end{align}
Here, it can be checked that $\breve{w}_{\pm}(0; s | \bm{w}^0) = \hat{p}_{\pm}(s
| \bm{w}^0)$ [Eq.~(\ref{e.fraction.p(s)})].  The equation
(\ref{e.app.w(u;s).general}) is a general expression of the forward recurrence
time PDF. In the following subsections, we derive more specific expressions for
the equilibrium and non-equilibrium ensembles.

\subsection {Equilibrium ensemble}

If both $\mu_+$ and $\mu_-$ exist, it follows form
Eq.~(\ref{e.app.w(u;s).general}) that
\begin{equation}
  \label{e.app.eq-init-ensenble}
  \lim_{s \to 0} s \breve{w}_{\pm}(u; s  | \bm{w}^0)
  =
  p_{\pm}^{\mathrm{eq}}\frac {1 - \hat{\rho}_{\pm} (u)}{\mu_{\pm} u}.
\end{equation}
This should be equivalent to the equilibrium ensemble
$\hat{w}^{\mathrm{eq}}_{\pm} (u)$. Therefore, the equilibrium ensemble is given
by $\hat{w}^{\mathrm{eq}}_{\pm} (u) =
p_{\pm}^{\mathrm{eq}}\hat{\rho}_{\pm}^{\mathrm{eq}}(u)$, where
$\hat{\rho}_{\pm}^{\mathrm{eq}}(u)$ is given by
Eq.~(\ref{e.init-deinsity-equilibrium}).  Replacing $\bm{w}^0$ in
Eq.~(\ref{e.app.w(u;s).general}) with this equilibrium density
$\bm{w}^{\mathrm{eq}}$, we have the forward recurrence time PDF starting from
the equilibrium ensemble:
\begin{equation}
  \label{e.app.w(u;s).eq}
  \breve{w}_{\pm}^{\mathrm{eq}}(u; s) := 
  \breve{w}_{\pm}(u; s  | \bm{w}^{\mathrm{eq}}) =
  p_{\pm}^{\mathrm{eq}}\frac {1 - \hat{\rho}_{\pm} (u)}{\mu_{\pm} us}.
\end{equation}
By double Laplace invasions of this equation, we have
\begin{equation}
  \label{e.app.w(tau;t').eq}
  w_{\pm}^{\mathrm{eq}}(\tau; t')
  =
  \frac {p_{\pm}^{\mathrm{eq}}}{\mu_{\pm}}
  \int_{\tau}^{\infty} \rho_{\pm} (\tau') d\tau'.
\end{equation}
Therefore, the equilibrium forward recurrence time PDF
${w}_{\pm}^{\mathrm{eq}}(\tau; t')$ does not depend on the elapsed time
$t'$. This is a natural consequence, considering the stationarity of the
equilibrium processes.

\subsection {Non-equilibrium ensemble}

For the non-equilibrium ensemble [Eq.~(\ref{noneq_initial})], we have
\begin{align}
  \breve{w}_{\pm}^{\mathrm{neq}}(u; s)
  &:= 
  \breve{w}_{\pm}(u; s | \bm{w}^{\mathrm{neq}}) \notag\\[0.0cm]
  \label{e.app.w(u;s).neq.0}
  &=
  \frac {p_{\pm}^0 + p_{\mp}^0 \hat{\rho}_{\mp}(s)}{1 - \hat{\rho} (s)}
  \frac {\hat{\rho}_{\pm} (u) - \hat{\rho}_{\pm} (s)} {s-u}.
\end{align}
With Eq.~(\ref{e.fraction.p(s).noneq}), the above equation can be rewritten as
\begin{equation}
  \label{e.app.w(u;s).neq}
  \breve{w}_{\pm}^{\mathrm{neq}}(u; s)
  =
  \frac {s\hat{p}^{\mathrm{neq}}_{\pm}(s)}{s-u}
  \frac {\hat{\rho}_{\pm} (u) - \hat{\rho}_{\pm} (s)} {1 - \hat{\rho}_{\pm}(s)}.
\end{equation}

\section {General transition probability}\label{s.joint.pdf.general}

To calculate the ACF $\left\langle D(t_1)D(t_2) \right\rangle$, here we derive
general transition probabilities $W_{hh'} (t_1, t_2|\bm{w}^0)$. Namely,
$W_{hh'}(t_1,t_2|\bm{w}^0)$ with $t_1 \leq t_2$ is a joint probability that the
state is $h$ at time $t_1$ and the state is $h'$ at time $t_2$, given that the
process starts with $w^0_{\pm}(\tau)$ at $t=0$.
%
In order to derive $W_{hh'}(t_1,t_2|\bm{w}^0)$, we first consider $W_{hh'}(\tau;
t_1|\bm{w}^0)$ which is also a joint probability of which the state is $h$ at
time $t_1$ and the state is $h'$ at time $t_1 + \tau$, given that the process
starts with $w^0_{\pm}(\tau)$ at $t=0$. Thus, $W_{hh'}(t_1,t_2|\bm{w}^0)$ and
$W_{hh'}(\tau; t_1|\bm{w}^0)$ are related as
\begin{equation}
  \label{e.W(t1,t2)=W(t2-t1;t1)}
  W_{hh'}(t_1,t_2|\bm{w}^0) =
  \begin{cases}
    W_{hh'}(t_2-t_1;t_1|\bm{w}^0) & (t_2\geq t_1), \\[.0cm]
    0                             & (t_2< t_1),
  \end{cases}
\end{equation}
and the Laplace transformation gives
\begin{equation}
  \label{e.W(s1,s2)=W(s1;s1+s2)}
  \breve{W}_{hh'}(s_1,s_2|\bm{w}^0) = \breve{W}_{hh'}(s_2;s_1+s_2|\bm{w}^0).
\end{equation}

Note that $W_{hh'} (\tau; t_1|\bm{w}^0)$ can be written with the transition
probability $W_{hh'} (\tau|\bm{w}^0)$ given in Eq.~(\ref{e.W(s).2}) as
\begin{equation}
W_{hh'} (\tau; t_1|\bm{w}^0)
=
W_{hh'} (\tau|\bm{w}'),
\end{equation}
where $w'_{\pm}(\tau) := w_{\pm}(\tau; t_1|\bm{w}^0)$ is the forward recurrence
time PDF analyzed in the previous section [Eq.~(\ref{e.app.w(u;s).general})].
Therefore, replacing $w^{0}_{\pm}(\tau)$ in Eq.~(\ref{e.W(s).2}) with
$w_{\pm}(\tau; t_1 |\bm{w}^0)$ and $p_{\pm}^0$ with $p_{\pm}(t_1|\bm{w}^0)$, we
have
\begin{align}
  \label{e.W(s;t')}
  \begin{split}
    \hat{W}_{\pm\pm} (u; t_1|\bm{w}^0) & = 
    \frac {p_{\pm}(t_1|\bm{w}^0)}{u} - \frac {\hat{w}_{\pm}(u; t_1|\bm{w}^0) }{u}
    \frac{1-\hat{\rho}_{\mp}(u)}{1- \hat{\rho}(u)},
    \\[.0cm]
    \hat{W}_{\pm\mp} (u; t_1|\bm{w}^0) & = 
    \frac {\hat{w}_{\pm}(u;t_1|\bm{w}^0) }{u}
    \frac{1-\hat{\rho}_{\mp}(u)}{1- \hat{\rho}(u)}.
    \\[-0.6cm]
  \end{split}
\end{align}
From Eqs.~(\ref{e.W(s1,s2)=W(s1;s1+s2)}) and (\ref{e.W(s;t')}), we have general
expressions for the transition probabilities:
\begin{align}
  \label{e.W(s1,s2)}
  \begin{split}
    \breve{W}_{\pm\pm} (s_1, s_2|\bm{w}^0) =& 
    \frac {\hat{p}_{\pm}(s_1+s_2|\bm{w}^0)}{s_2}
    \\[0.0cm]
    -& \frac {\breve{w}_{\pm}(s_2; s_1+s_2|\bm{w}^0) }{s_2}
    \frac{1-\hat{\rho}_{\mp}(s_2)}{1- \hat{\rho}(s_2)},
    \\[.0cm]
    \breve{W}_{\pm\mp} (s_1, s_2|\bm{w}^0) =& 
    \frac {\breve{w}_{\pm}(s_2;s_1+s_2|\bm{w}^0) }{s_2}
    \frac{1-\hat{\rho}_{\mp}(s_2)}{1- \hat{\rho}(s_2)}.
  \end{split}
\end{align}

For the equilibrium ensembles, let us define $\breve{W}_{hh'}^{\mathrm{eq}}
(s_1, s_2) := \breve{W}_{hh'} (s_1, s_2|\bm{w}^{\mathrm{eq}})$. Then, from
Eqs.~(\ref{e.fraction.p(s).eq}), (\ref{e.app.w(u;s).eq}) and (\ref{e.W(s1,s2)}),
we obtain
\begin{align}
W_{hh'}^{\mathrm{eq}} (t_1, t_2) =
  \begin{cases}
    W_{hh'}^{\mathrm{eq}} (t_2- t_1) & (t_2\geq t_1), \\[.0cm]
    0                                & (t_2< t_1),
  \end{cases}
\end{align}
where $W_{hh'}^{\mathrm{eq}} (t)$ is given by Eq.~(\ref{e.W(s).eq}). This
relation implies a famous property of equilibrium processes; namely, any
two-time correlation functions $\langle A(t_1)B(t_2) \rangle$ in equilibrium
processes depend only on the time lag $t_2-t_1$.

For the non-equilibrium ensembles, we define $\breve{W}_{hh'}^{\mathrm{neq}} (s_1,
s_2) := \breve{W}_{hh'} (s_1, s_2|\bm{w}^{\mathrm{neq}})$,
and then simply rewrite Eq.~(\ref{e.W(s1,s2)}) as
\begin{align}
  \label{e.W(s1,s2).neq}
  \begin{split}
    \breve{W}_{\pm\pm}^{\mathrm{neq}} (s_1, s_2) & = 
    \frac {\hat{p}_{\pm}^{\mathrm{neq}}(s_1 + s_2)}{s_2} -
    \frac {\breve{w}_{\pm}^{\mathrm{neq}}(s_2; s_1 + s_2) }{s_2}
    \frac{1-\hat{\rho}_{\mp}(s_2)}{1- \hat{\rho}(s_2)},
    \\[.0cm]
    \breve{W}_{\pm\mp}^{\mathrm{neq}} (s_1,s_2) & = 
    \frac {\breve{w}_{\pm}^{\mathrm{neq}}(s_2;s_1 + s_2) }{s_2}
    \frac{1-\hat{\rho}_{\mp}(s_2)}{1- \hat{\rho}(s_2)}.\\[-.7cm]
  \end{split}
\end{align}

\section {Double Laplace inversions}\label{s.double.laplace.inversion}
\begin{figure}[]
  \centerline{\includegraphics[width=5.3cm]{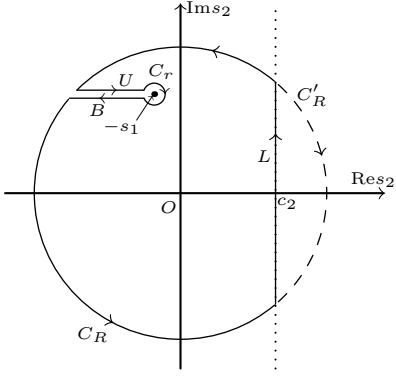}}
  \caption{\label{f.complex-path-case1} Integration path on the complex $s_2$
    plane. $s_2 = -s_1$ is a branch point, and we take a branch cut on $s_2 =
    -s_1 - b$ with a real number $b \in (0, \infty)$.}
\end{figure}

Here, we briefly outline the double Laplace inversions of
Eq.~(\ref{e.<dD(s1)dD(s2)>/s1s2.case(1-1)}). First, we consider
\begin{align}
\mathcal{L}^{-2}
&\left[
\frac {s_1^{\alpha_-}}{s_1^2s_2^2(s_1+s_2)^{\alpha_+}}
\right](t_1, t_2)
:=
\notag\\[0.0cm]
\label{e.app.double-laplace-inversion.case(1).1}
&
\int_{c_1-i\infty}^{c_1+i\infty} \frac {ds_1}{2 \pi i}
\int_{c_2-i\infty}^{c_2+i\infty} \frac {ds_2}{2 \pi i}
\frac {e^{s_1t_1}e^{s_2t_2} s_1^{\alpha_-}}{s_1^2s_2^2(s_1+s_2)^{\alpha_+}},
\end{align}
where $\alpha_{\pm} \in (0,1)$, but these parameter ranges can be extended (see
below).  Differentiating the above equation two times with respect to $t_2$, we
obtain
\begin{align}
  \frac {\partial^2 }{\partial t_2^2}
  &\mathcal{L}^{-2}
  \left[
  \frac {s_1^{\alpha_-}}{s_1^2s_2^2(s_1+s_2)^{\alpha_+}}
  \right](t_1, t_2)
  =
  \notag\\[0.0cm]
\label{e.app.double-laplace-inversion.case(1).2}
  &
  \int_{c_1-i\infty}^{c_1+i\infty}\frac {ds_1}{2 \pi i}
  \frac {e^{s_1t_1}}{s_1^{2-\alpha_-}}
  \int_{c_2-i\infty}^{c_2+i\infty}\frac {ds_2}{2 \pi i}
  \frac {e^{s_2t_2}}{(s_1+s_2)^{\alpha_+}}.
\end{align}
The integral in terms of $s_2$ can be calculated in a standard way
\cite{schiff99}; we consider the complex integration path $C_R+U+C_r+B+L$
displayed in Fig.~\ref{f.complex-path-case1}. The inside of this complex path is
analytic, and thus the integration along the path becomes 0 due to the Cauchy's
theorem. Also, it can be checked that the integrations along $C_R$ and $C_r$
tend to 0, as $R \to \infty$ and $r \to 0$, respectively ($R$ and $r$ are
radii of $C_R$ and $ C_r$).
Thus, the integration paths only along the branch cut, $U$ and $B$, contribute
to the integration along $L$:
\begin{align}
  \lim_{R \to \infty}
  \int_{L}\frac {ds_2}{2 \pi i}
  \frac {e^{s_2t_2}}{(s_1+s_2)^{\alpha_+}}
  &=
  -\lim_{R \to \infty}
  \int_{U+B}\frac {ds_2}{2 \pi i}
  \frac {e^{s_2t_2}}{(s_1+s_2)^{\alpha_+}}
  \notag\\[0.0cm]
\label{e.app.double-laplace-inversion.case(1).3}
  &=
  e^{-s_1t_2} \frac {t_2^{\alpha_+-1}}{\Gamma(\alpha_+)}.
\end{align}
Putting this equation into Eq.~(\ref{e.app.double-laplace-inversion.case(1).2}),
and integrating in terms of $s_1$, we have
\begin{equation}
\label{e.app.double-laplace-inversion.case(1).4}
  \frac {\partial^2 }{\partial t_2^2}
  \mathcal{L}^{-2}\!
  \left[
  \frac {s_1^{\alpha_-}}{s_1^2s_2^2(s_1+s_2)^{\alpha_+}}
  \right]\!(t_1, t_2)
  =
  \frac
  {(t_1-t_2)^{1-\alpha_-} t_2^{\alpha_+-1}}
  {\Gamma(\alpha_+) \Gamma(2-\alpha_-)}.
\end{equation}
%
Integrating this equation (from 0 to $t_2$) two times with respect to $t_2$, and
setting $t_1=t_2=t$, we obtain
\begin{equation}
\label{e.app.double-laplace-inversion.case(1).5}
  \mathcal{L}^{-2}\!
  \left[
  \frac {s_1^{\alpha_-}}{s_1^2s_2^2(s_1+s_2)^{\alpha_+}}
  \right]\!(t, t)
  =
  \frac {(2-\alpha_-)t^{2 + \alpha_+-\alpha_-}}{\Gamma(3+\alpha_+ -\alpha_-)},\!
\end{equation}
where the terms of the orders $O(t_2^0)$ and $O(t_2^1)$, appearing on the left
hand side when integrating with respect to $t_2$, vanish. This is because, if we
take a closed integration path $L+C_R'$ depicted in
Fig.~\ref{f.complex-path-case1} for these terms, the integration along $L+C_R'$
becomes 0 by virtue of the Cauchy's theorem, whereas the integration along
$C_R'$ also vanishes as $R\to \infty$ since the exponential factor $e^{s_2t_2}$
is absent in these terms. It follows that the integrations along $L$ of the
terms $O(t_2^0)$ and $O(t_2^1)$ converge to 0 as $R \to \infty$.


Next, in Eq.~(\ref{e.app.double-laplace-inversion.case(1).5}), substituting 0
for $\alpha_-$ and then substituting $\alpha_- - \alpha_+$ for $-\alpha_+$ [this
procedure is justified, because, with respect to both $\alpha_+$ and $\alpha_-$,
the right hand side of Eq.~(\ref{e.app.double-laplace-inversion.case(1).5}) can
be analytically continued into the whole complex planes except poles thanks to
the analyticity of the gamma function], we obtain
\begin{equation}
\label{e.app.double-laplace-inversion.case(1).6}
  \mathcal{L}^{-2}
  \left[
  \frac {(s_1+s_2)^{\alpha_- - \alpha_+}}{s_1^2s_2^2}
  \right](t, t)
  =
  \frac {2 t^{2+\alpha_+- \alpha_-}}{\Gamma(3+\alpha_+-\alpha_-)}.
\end{equation}
Using Eqs.~(\ref{e.app.double-laplace-inversion.case(1).5}) and
(\ref{e.app.double-laplace-inversion.case(1).6}), we have
Eq.~(\ref{e.RSD(t)-case(1-1)}) from Eq.~(\ref{e.<dD(s1)dD(s2)>/s1s2.case(1-1)}).
Double Laplace inversions of Eqs.~(\ref{e.<dD(s1)dD(s2)>/s1s2.case(2-1)}) and
(\ref{e.<dD(s1)dD(s2)>/s1s2.case(3a-2)}) can be carried out in a similarly way.

\section {Non-Gaussian parameter}\label{s.non-gaussian}

A non-Gaussian parameter of the displacement vector $\delta\bm{r}(t)$ is defined
by \cite{kegel00, arbe02, ernst14, cherstvy14}
\begin{equation}
  A(t) := \frac {n}{n+2}
  \frac
  {\left\langle \delta \bm{r}^4(t) \right\rangle}
  {\left\langle \delta \bm{r}^2(t) \right\rangle^2} - 1,
\end{equation}
where $n$ is the spatial dimension. This parameter $A(t)$ is 0 if
$\delta\bm{r}(t)$ follows the multi-dimensional Gaussian distribution. For the
two-state LEFD [Eq.~(\ref{e.lefd})], we have
\begin{equation}
  \left\langle \delta \bm{r}^4(t) \right\rangle
  = 4 n(n+2) \int_{0}^{t}dt' \int_{0}^{t}dt'' \left\langle D(t')D(t'') \right\rangle,
\end{equation}
where we used the Wick's theorem [Eq.~(\ref{e.wick})]. From this equation and
Eq.~(\ref{e.emsd}), we have
\begin{equation}
  A(t) =
  \frac
  {\displaystyle\int_{0}^{t}dt' \int_{0}^{t}dt''
    \left\langle \delta D(t')\delta D(t'') \right\rangle}
  {\displaystyle\int_{0}^{t}dt' \int_{0}^{t}dt''
    \left\langle D(t')\right\rangle \left\langle D(t'') \right\rangle},
\end{equation}
which is equivalent to the excess RSD, $\Sigma^2_{\mathrm{ex}} (t | \bm{w}^0)$
[Eq.~(\ref{e.RSD(t).sigma_ex})]. Thus, the RSD analysis is a way to extract
non-Gaussianity from trajectory data.
Note, however, that the excess RSD is not equivalent to $A(t)$ for the
anisotropic systems or systems with orientational correlations \cite{uneyama15}.

\section {Simulation setup}\label{sec:app.simulation}

In numerical simulations, we used ideal sojourn time PDFs defined as
\begin{equation}
  \rho_{\pm} (\tau) = \frac {\alpha_{\pm} \tau_0^{\alpha_{\pm}}}{\tau^{1+\alpha_{\pm}}},
  \qquad
  \tau \in [\tau_0, \infty),
\end{equation}
where $\tau_0$ is a cutoff time for short trap times; we assumed the same cutoff
time for both $\rho_+(\tau)$ and $\rho_-(\tau)$. Note that $\hat{\rho}_{\pm}(s)$
can be written in the forms of Eqs.~(\ref{e.rho(s).asymptotic.alpha<1}) or
(\ref{e.rho(s).asymptotic.alpha>1}).
In fact, if $\alpha_{\pm} \in (0,1)$, the term $O(s)$ in
Eq.~(\ref{e.rho(s).asymptotic.alpha<1}) is given by $\alpha_{\pm}\tau_0
s/{(1-\alpha_{\pm})}$.
On the other hand, if $\alpha_{\pm} \in (1,2)$, then $\mu_{\pm} =
\alpha_{\pm}\tau_0/(\alpha_{\pm}-1)$, and the term $O(s^2)$ in
Eq.~(\ref{e.rho(s).asymptotic.alpha>1}) is given by $- {\alpha_{\pm}\tau_0^2
  s^2}/{2(2-\alpha_{\pm})}$.  Moreover, $a_{\pm}$ is given by $a_{\pm}=
\tau_0|\Gamma(1-\alpha_{\pm})|$ for $\alpha_{\pm} \in (0,2)$.

Then, the Langevin equation given in Eq.~(\ref{e.lefd}) or
\begin{equation}
   d\bm{r}(t) = \sqrt{2 D(t) dt}\, \bm{\xi}(t),
\end{equation}
can be transformed into dimensionless form with
\begin{equation}
  \tilde{\bm{r}}(t) = \frac {\bm{r}(t)}{\sqrt{D_+ \tau_0}}, \qquad
  \tilde{t} = \frac {t}{\tau_0}.
\end{equation}
The remaining system parameters are $\alpha_{\pm}$ and the ratio $D_-/D_+$. In
simulations, we set $D_-/D_+ = 0.02$; this is because, in \cite{parry14},
diffusion in bacterial cells has been found to have a fast and a slow states
with different DCs and the ratio of these two DCs are about 50 (They reported
that a histogram of a radius of gyration $R_g$, which is proportional to the
square root of the DC, has two peaks. $R_g$ of one peak is about 7 times bigger
than that of the other. This means that the ratio of the two DCs is about $7^2$,
i.e., $D_+/D_- \approx 50$.).

Moreover, to simulate equilibrium processes, we have to generate initial
ensembles which follow the first sojourn time PDFs $\rho_{\pm}^{\mathrm{eq}}(t)$
[Eq.~(\ref{e.init-deinsity-equilibrium})]. This can be achieved with a method
presented in \cite{miyaguchi13}. As a scheme of numerical integration of the
Langevin equation, the Euler method is used \cite{kloeden11}.


\bibliography{paper}

\end {document}